\documentclass[12pt,preprint]{aastex}
\usepackage{natbib,amssymb}
\newcommand{\mic}{ $\mu$m }
\bibpunct{(}{)}{;}{a}{}{,}
\usepackage{graphics,graphicx,subfigure}

\begin{document}

\title{Spitzer IRS Observations of Class I/II Objects in Taurus: \\ Composition and Thermal History of the Circumstellar Ices}
\shorttitle{Characteristics of Ices in Taurus Class I/II Objects}
\shortauthors{G.~Zasowski et al.}

\author{G.~Zasowski\altaffilmark{1},
F.~Kemper\altaffilmark{2,1}, Dan~M.~Watson\altaffilmark{3},
E.~Furlan\altaffilmark{4,5}, C.J.~Bohac\altaffilmark{3},
C.~Hull\altaffilmark{1}, and
J.D.~Green\altaffilmark{3}} 

\altaffiltext{1}{Department of Astronomy, PO Box 400325, University of Virginia, Charlottesville, VA 22904-4325: gailis@virginia.edu, chat.hull@gmail.com}
\altaffiltext{2}{Jodrell Bank Centre for Astrophysics, University of Manchester, Manchester, M13 9PL, UK: Ciska.Kemper@manchester.ac.uk}
\altaffiltext{3}{Department of Physics and Astronomy, University of Rochester, Bausch \& Lomb Hall, PO Box 270171, 500 Wilson Blvd, Rochester, NY 14627-0171: dmw@pas.rochester.edu, joel@pas.rochester.edu, urgradcb@gmail.com}
\altaffiltext{4}{NASA Astrobiology Institute and Department of Physics and Astronomy, UCLA, 430 Portola Plaza, Los Angeles, CA 90095: furlan@astro.ucla.edu}
\altaffiltext{5}{Current address: Jet Propulsion Laboratory, Caltech, Mail Stop 264-767, 4800 Oak Grove Drive, Pasadena, CA 91109}

\begin{abstract}
We present observations of Taurus-Auriga Class I/II protostars obtained with the Spitzer InfraRed Spectrograph.
Detailed spectral fits
to the 6 and 15.2 micron ice features are made, using publicly-available laboratory data, to constrain the
molecular composition, abundances, and levels of thermal processing along the lines of sight.  
We provide an inventory of the molecular environments observed, which have an average composition dominated
by water ice with $\sim$12\% CO$_2$ (abundance relative to H$_2$O), $\gtrsim$2-9\% CH$_3$OH, $\sim$14\% NH$_3$, $\sim$3\% CH$_4$,
$\sim$2\% H$_2$CO, $\sim$0.6\% HCOOH, and $\sim$0.5\% SO$_2$.  We find CO$_2$/H$_2$O ratios nearly equivalent to those observed
in cold clouds and lines of sight toward the galactic center.  The unidentified 6.8 micron profiles
vary from source to source, and it is shown to be likely that even combinations of the most common candidates
(NH$_4^+$ and CH$_3$OH) are inadequate to explain the feature fully.  
We discuss correlations among SED spectral indices, abundance ratios, and thermally-processed ice fractions and their implications for CO$_2$ formation and evolution.
Comparison of our spectral fits
with cold molecular cloud sight-lines indicates abundant prestellar ice environments made even richer by
the radiative effects of protostars.  Our results add additional constraints and a finer level of detail to current
full-scale models of protostellar and protoplanetary systems.
\end{abstract}



\section{INTRODUCTION}

The chemical and thermal evolution of protostellar environments affects the spectral appearance
of ices observed in the sequential stages of pre-main-sequence evolution
\citep[e.g.][]{WGH_82_protostars,GWE_99_molcloud,vD_06_PNASarticle}. The
presence of the ice absorption features is often easily recognized, with the most common ones
in the mid-infrared wavelength range found at
6.0, 6.8, and 15.2 $\mu$m, but the exact feature shapes depend strongly on
the composition of the ice as determined by the initial condensation
and subsequent thermal processing and UV irradiation. 
Mixing ices can cause wavelength shifts in the resonances of the individual components within the ice matrix
\citep[e.g.][]{EKS_99_laboratory,PB_00_icymixtures,OFB_07_effectco2onh2o}, and
differences in lattice structure caused by either radiation damage or thermal processing are spectroscopically identifiable
\citep[e.g.][]{HTG_81_waterice,MH_92_waterice,MSR_98_waterice,KK_90_UVamorphization}.  

The initial reservoir of star formation material, the diffuse ISM, is extremely ice-poor \citep[e.g.][]{Gillett_75_DISMice,Whittet_97_DISMdust}.  
It is not dense enough to shield its dust grains, upon which ices condense \citep[e.g.][]{TielensAlla_87_ISMdust}, from incident destructive photons.  Only when cold
molecular clouds of high density ($n_H$ $\gtrsim$ 10$^4$ cm$^{-3}$) are formed can interstellar ices survive and be observed.  
In the Taurus molecular cloud, for example, 
a well-known seat of starbirth, ices are observed towards background stars even in quiescent regions unaffected by the newly-formed stars 
\citep[e.g.][]{BMG_05_co2fieldstars,KBP_05_backgroundstars,WSB_07_quiescentICM}.  These studies reveal
absorptions attributable to H$_2$O, CO, and CO$_2$, as well as some less-constrained absorptions (e.g. the still unidentified band at 6.8 $\mu$m) 
due to presumably more complex species.  
The ice bands observed towards these field stars are smooth, indicating amorphous and unprocessed compounds, and low upper limits can be placed on some species
created by UV radiation \citep[e.g. ``XCN'';][]{GWB_04_ice}.  These observations support 
the theory that molecular cloud ice chemistry is dominated by efficient cold grain surface reactions.  The presence of larger complex molecules \citep[such as HCOOH, 
which can be formed at low temperatures;][]{Keane_01_makehcooh} demonstrates the potential complexity of these reactions and 
indicates that protostars are born into a pre-enriched chemical environment.

Surveys of deeply embedded young stellar objects (YSOs) 
reveal a limited range of thermal processing, indicative of differences in YSO environment
and temperature \citep[e.g.][]{GWE_99_molcloud,GWB_04_ice}.
Studies of less-extinguished protostars \citep[Class I in the classification of][]{Lada_87_starformation} show evidence 
of enhanced complex molecule formation (H$_2$CO, XCN, OCS, PAHs [polycyclic aromatic hydrocarbons]), 
ice crystallization, further thermal annealing, and ice matrix segregation 
\citep[see e.g.][]{ACA_03_isocam,KTB_01_5-8umregion,EDD_98_segregation,WKC_04_taurus,GWB_04_ice,BPL_04_lowmass}.

Finally, the least embedded Class II objects have seemingly accreted or dissipated most of their 
cold natal envelopes and have heated the outer layers of their dust disks above the 
sublimation temperature of astrophysical ices.  These objects tend to show silicate dust \emph{emission}, PAH emissions, and weak or 
nonexistant ice signatures \citep[e.g.][]{Forrest_04_ttauri,Furlan_06_ttauristars}, 
though some Class II edge-on systems have ice absorptions arising from the cold disk midplane \citep[e.g.][]{Pontopp_05_diskices,Furlan_08_classImodel}.

Differences in chemical and physical structure exist between YSOs with masses either above or below a few M$_\sun$
\citep{vanDishoeck_03_highlowmassYSOs}.  Some of these
dissimilarities may be attributed to observational biases (low-mass YSOs are generally closer and can be better resolved), but others are likely to be
dependent on the intrinsic nature of the object (e.g. the role of shocks, core temperature, and infall timescales).  
For example, low-mass YSO studies claim higher abundances of CO$_2$, relative to H$_2$O \citep{BPL_04_lowmass,Pontop_08_co2inlowmass} 
and amorphous silicate dust \citep{WKC_04_taurus}, than surveys of more massive YSOs.
This paper focuses on the relatively underexplored categories of low-mass Class I protostars and Class I/II transition objects 
(henceforth both groups will be collectively referred to as Class I/II objects)
with ice absorptions by analyzing mid-infrared spectroscopy of 16 Taurus-Auriga YSOs \citep[$<$1-3 Myr old;][]{White_01_TMCage}, observed with
the InfraRed Spectrograph \citep[IRS;][]{HRV_04_IRS} on board the Spitzer Space Telescope \citep{WRL_04_Spitzer}.
Our sample overlaps slightly with that of the extensive recent work by the {\it c2d} team \citep[e.g.][]{Boogert_08_lowmassices,Pontop_08_co2inlowmass,Oberg_08_ch4}, 
and we indicate differences in our approach and findings throughout this paper.
This study helps complete our view of the ice evolutionary sequence from cold molecular clouds through
warmer environments with protoplanetary disks.  These low-mass, possibly early solar analogs are essential pieces to the puzzle of solar system formation.

Section~\ref{sec:obs_reduction} contains a short description of the sample, observations, and data reduction.  In Section~\ref{sec:methods} we describe 
how the ice features are identified, isolated, and analyzed, and in Section~\ref{sec:trends} we look at trends seen in the feature optical depths.
In Section~\ref{sec:features} we derive the composition of the carriers of the primary ice absorptions and evaluate 
candidates for some of the less constrained features.  In Section~\ref{sec:discussion} we discuss composition and spatial distribution of the ices and place
our observations in the context of pre-main-sequence evolution.

\section{OBSERVATIONS} \label{sec:obs_reduction}

\subsection{Target Sample Description}

The Taurus-Auriga dark cloud is a nearby \citep[d $\sim$ 140 pc;][]{Bertout_99_TMCdistance}, low density \citep[$n_*$ $\sim$ 1-10 pc$^{-3}$;][]{Gomez_93_TMCdensity} 
star forming region containing predominantly young, low-mass protostars.  
It provides an opportunity to study closely low-mass star formation unaffected by the shocks and strong UV radiation associated with more massive protostars.  
Based on IRAS 25 and 60\mic photometry, an
extensive sample of Class I/II objects was selected from the list by \citet{KenHart_95_YSOlist} and observed with Spitzer's IRS (Section~\ref{sec:obsprocedure}). 
From this sample we chose for analysis the subset of 12 Class I and 4 Class II objects \citep[as classified in][]{KenHart_95_YSOlist} 
with distinct ice features in regions of unsaturated, well-calibrated data.  
The IRAS spectral energy distribution (SED) points used in the original selection are insufficient to determine system inclination, 
but modeling by \citet{Furlan_08_classImodel} indicates that 
our smaller sample spans a wide range of disk inclination angles, from nearly edge-on to face-on systems.

For comparison, we include a secondary sample comprising observations of Taurus and Serpens cloud field stars---Elias 16, Elias 13, and CK 2---as well 
as of the Class 0 protostars CepE-MM and L1448 IRS 2 (AOR IDs 3623680, 3620352).
We also include data from the line of sight towards the W3(OH) star formation region, which probes part of the foreground W3 molecular cloud (AOR ID 3566336).

\subsection{Observational Procedure and Data Reduction} \label{sec:obsprocedure}
Our sample was observed with Spitzer's IRS as part of the \emph{IRS Disks} project (PI: J.~Houck \& D.~Watson; PID 2)
during guaranteed time awarded to the IRS instrument team (2004 February-March).  All of our targets were observed using the full 5--37 $\mu$m
wavelength coverage, combining either the Short Low and Long Low (SL, LL; 5.2-14 $\mu$m, 14-38 $\mu$m; $\lambda$/$\Delta\lambda\,\sim$ 90)
modules on the IRS, or the SL with the Short High and
Long High (SH, LH; 10-19 $\mu$m, 19-37 $\mu$m; $\lambda$/$\Delta\lambda\,\sim$ 600) modules, depending on the expected flux levels. 

Most (12) of the objects were observed using a small spectral map ($3\times
2$) rather than a single pointing; this strategy avoids the need for pointing peak-ups and allows, in principle, for a better reconstruction of
the spectrum in cases of small mispointings.
The three-step sequence was carried out in the dispersion direction, with 
the steps separated by three quarters (for the SL module) or one half (for the other modules) of the slit width.
The two-step sequence corresponds to the normal IRS nod distance of one-third the slit length.
The other 4 objects were observed in the IRS staring mode, using only the two nods along the slit.
Observational details of all our Class I/II targets are listed in Table~\ref{tab:obs}.

We analyzed data from pipeline version S13, using
the {\sc smart} package \citep{HDH_04_smart}.  The standard {\it IRS Disks} reduction procedures, as well as any spectra requiring special treatment, are 
described in \citet{Furlan_06_ttauristars} and \citet{Furlan_08_classImodel}.

The uncertainty in each wavelength bin is taken to be half of the nod difference in flux level, except where this value is smaller than 1\% of the flux.  In those cases,
the uncertainties are taken to be 1\% of the flux in order to prevent underestimation.  We estimate an absolute spectrophotometric accuracy of 5-10\%.

For the field stars, we use the previously published data for 
Elias 16 \citep{BMG_05_co2fieldstars} and Elias 13 and CK~2 \citep{KBP_05_backgroundstars}, without re-reducing the data.

\subsection{IRS Spectra Description}

\begin{figure}
 \begin{center}
 \includegraphics[width=0.9\textwidth]{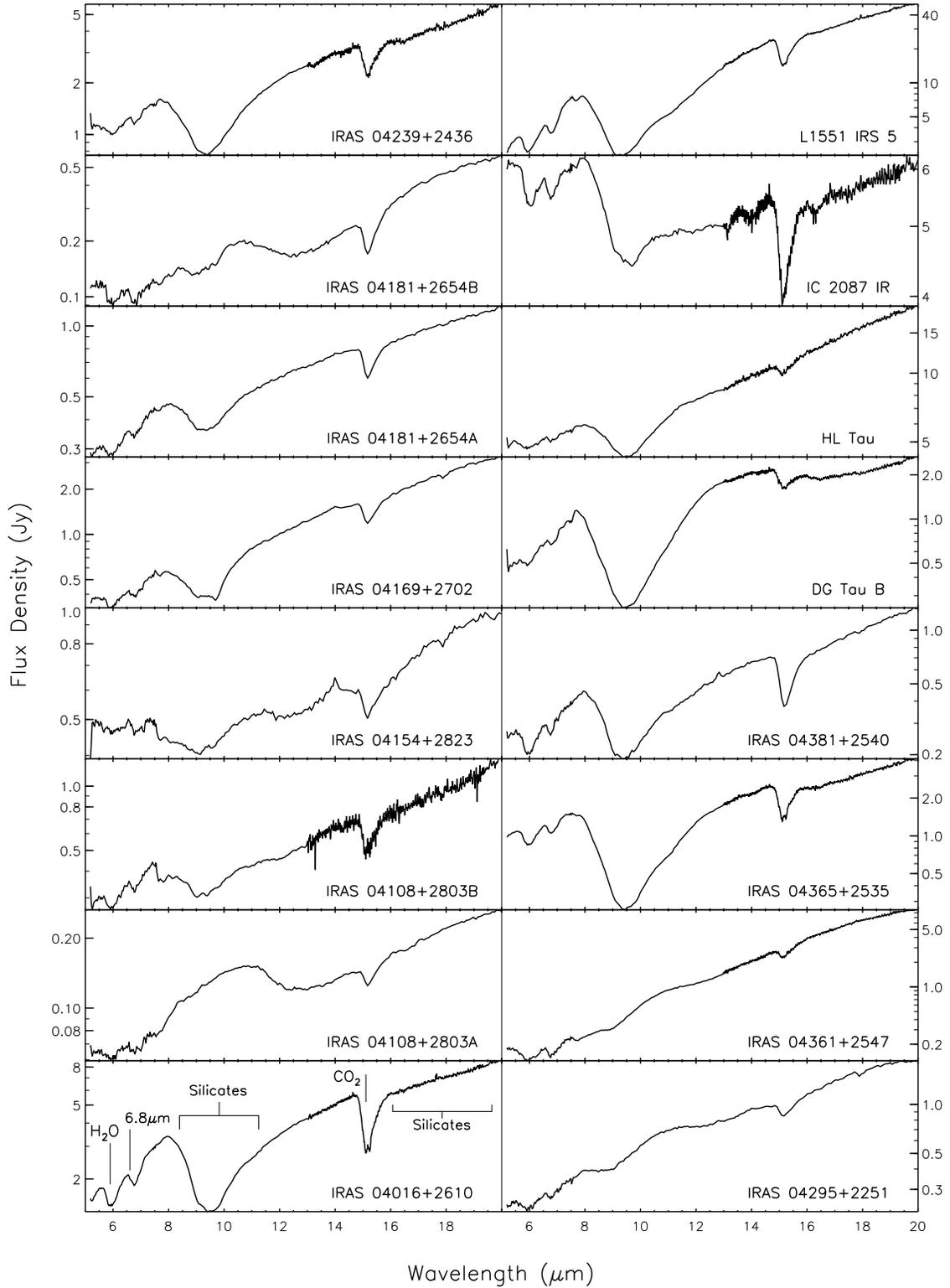}
 \end{center}
 \caption{Our sample IRS spectra.  The lower left object has been marked with the locations and primary carriers of the most prominent absorption features.  Note
that the 9.7\mic silicate feature may also appear partly or wholly in emission.}
 \label{fig:fullspectra}
\end{figure}

In Figure~\ref{fig:fullspectra}, we show the IRS spectra of the 16 Class I/II stars in our sample.
The YSO mid-infrared spectra are dominated by broad solid state resonances, particularly those at 6.0, 6.8, 9.7, and
15.2 $\mu$m.  Additional, but less pronounced, features are visible in some spectra at 7.7\mic and as a broad band centered at 18 $\mu$m.  Silicates are responsible
for the broad 9.7 and 18\mic features, which can occur in either absorption (e.g. L1551 IRS 5), emission (e.g. IRAS 04108+2803A), or both.

The 6.0\mic feature is primarily due to the O-H bending mode of water ice, while the 6.8\mic feature remains largely
unidentified, with CH$_3$OH and NH$_4^+$ (among others) as suggested carriers.  The 7.7\mic absorption is generally attributed to solid CH$_4$, and
the 15.2\mic band is due to the O-C-O bending mode of solid CO$_2$.
A complete inventory of absorption features identified in our spectra is contained in Table~\ref{tab:features}.

\begin{figure}
 \begin{center}
 \includegraphics[width=0.7\textwidth]{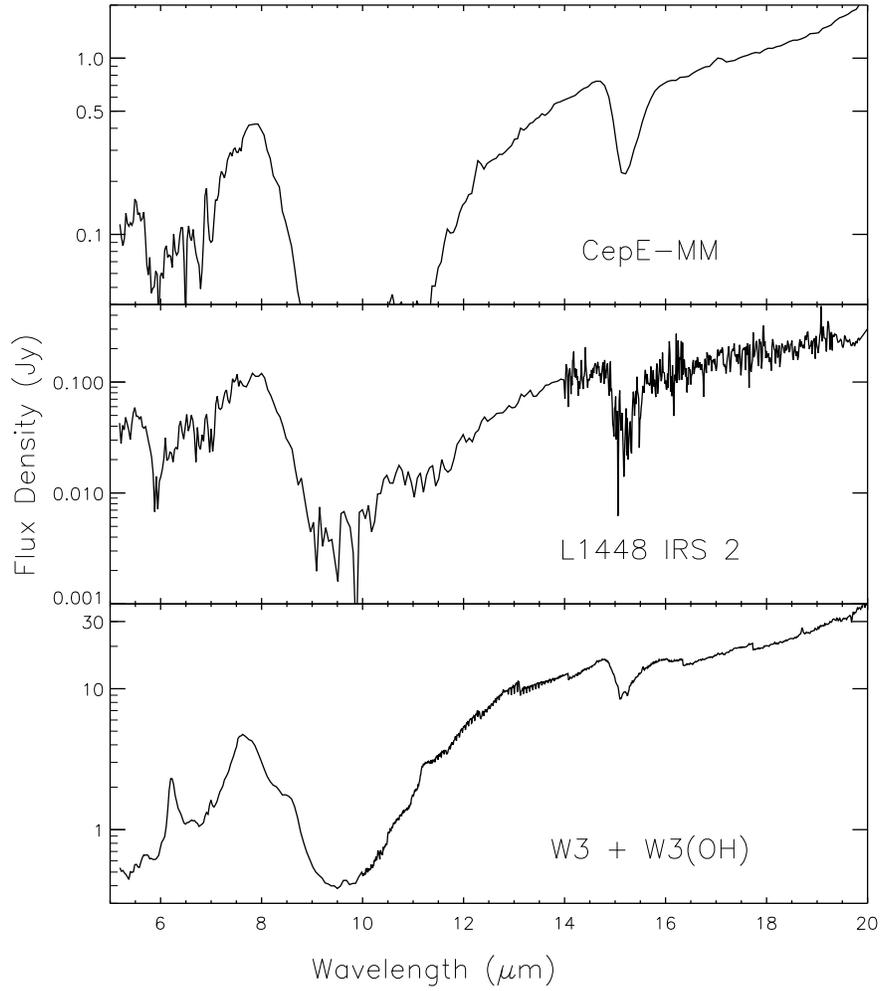}
 \end{center}
 \caption{IRS spectra of the Class 0 protostars CepE-MM (in Cepheus) and L1448 LRS 2 (in Perseus), and the line of sight to W3(OH) through the W3 molecular cloud.
The 9.7\mic silicate absorption feature in CepE-MM is saturated.}
 \label{fig:otherobjects}
\end{figure}

Spectra for W3(OH) and the Class 0 objects (Figure~\ref{fig:otherobjects}) contain many of the same features.

\section{METHODS} \label{sec:methods}

\subsection{Isolating the IR Features and Determining Optical Depths} \label{sec:isolate_features}

\subsubsection{YSO Spectra} \label{sec:isolate_features_irs}

The optical depth, $\tau$, and shape of the individual absorption features
under consideration are extracted from the astronomical spectra
using a mathematical continuum fit.  
We make a cubic spline fit to the $\sim$5-20\mic continuum of each
spectrum using the following relatively feature-free regions: bluewards of the 6\mic H$_2$O feature; between the unidentified 6.8\mic feature
and the onset of the 9.7\mic silicate band; between the red wing of the silicate band ($\sim$14.5 $\mu$m) and the 15.2\mic CO$_2$ feature;
and redwards of the CO$_2$ feature---while ensuring that
none of the feature wings are included in the continuum.
We calculate the optical depth in the spectral features using $F_\lambda=F_{\lambda,cont}e^{-\tau_\lambda}$.  

Because the 9.7\mic silicate feature is highly complex, and
detailed modeling of the silicate mineralogy is beyond the scope of this paper, accurate extraction of ice features blended with the 9.7\mic silicate complex 
(e.g. those of CH$_3$OH, NH$_3$, and H$_2$O) is impractical.
However, the 11-13\mic H$_2$O libration band and
the 18\mic absorption feature in particularly silicate-rich spectra are much broader and shallower than the 15.2\mic ice absorption they overlap, so 
we can treat these shapes as pseudo-continua in order to isolate the CO$_2$ ice absorption.

\begin{figure}
 \begin{center}
 \includegraphics[width=0.8\textwidth]{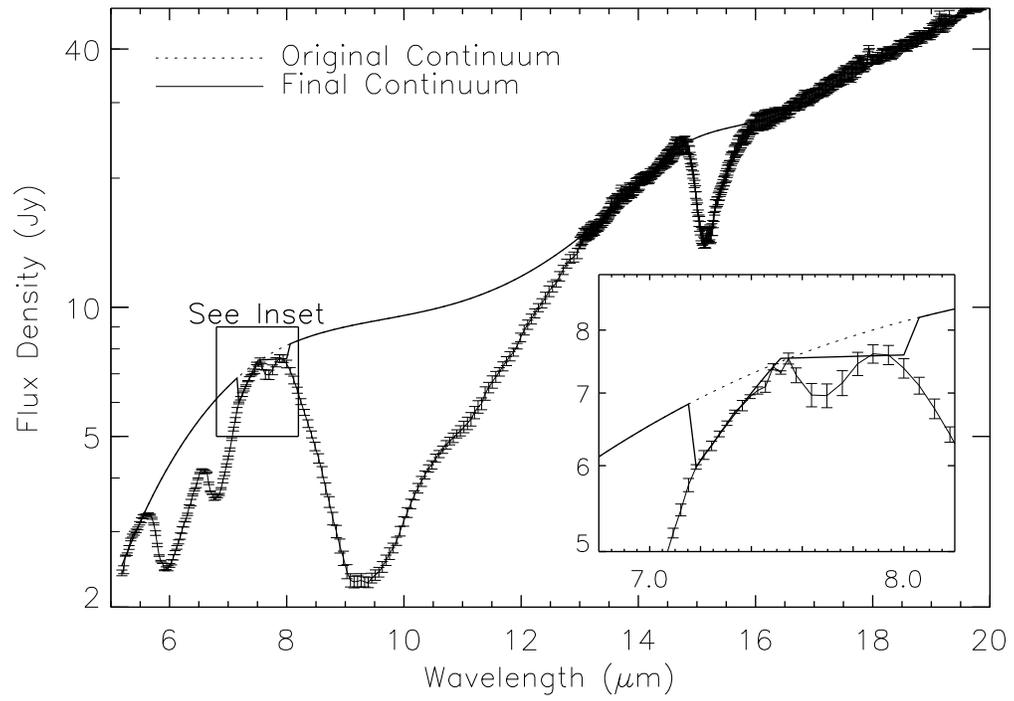}
 \end{center}
 \caption{Example of the applied continuum determination using the spectrum of L1551 IRS 5.  The inset shows the additional continuum imposed on the 7-8\mic range.}
 \label{fig:continuum-ex}
\end{figure}

Additional steps are needed to extract small features in the 7-8\mic range.  The wide variation in 6.8\mic and
silicate absorption profile wings requires that each spectrum be treated individually.
For objects not affected by the silicate feature bluewards of 8 $\mu$m, a straight-line continuum ($\sim$7.1-8 $\mu$m) is imposed to isolate any features.  
For objects affected by both the 6.8 and the 9.7\mic features, two straight-line continua are imposed: 
one from $\sim$7.1-7.5\mic and one from $\sim$7.5-8 $\mu$m (Figure~\ref{fig:continuum-ex}).

\subsubsection{Laboratory Spectra} \label{sec:isolate_features_lab}

We use laboratory data from the Leiden University ice analog databases\footnote{http://www.strw.leidenuniv.nl/$\sim$lab/databases/databases.htm}
\citep{GSG_95_labsim,Gerakines_96_puredb,EBG_96_coco2iso,EKS_99_laboratory,vBGF_06_coco2mixes} to identify the composition and apparent temperatures of
the ices seen in our targets. In order to provide a direct comparison between
the feature optical depths measured in the laboratory and those observed in the
astronomical spectra, we removed a straight-line continuum from the laboratory data.  For
the wavelength ranges of interest, optical depths calculated using a straight-line continuum differ, on
average, by less than half of a percent from those calculated using a more complicated cubic spline
one.  The wavelengths of the baseline ends are
generally equal to the continuum points used in the astronomical
spectra's spline fitting, with an occasional
slight shift to ensure feature wings are excluded from the continuum.  In Figure~\ref{fig:h2oco2_temp} we show
the range of abundances contained in the Leiden databases' non-irradiated H$_2$O:CH$_3$OH:CO$_2$
mixtures (T=10-185K, all circles); note that this plot does not represent the fraction of the
database comprising other molecules (e.g. CH$_4$ and H$_2$OH) or UV-irradiated mixtures.

The temperatures given in this paper, except where otherwise noted, refer to laboratory conditions.  Interstellar temperature barriers for crystallization, sublimation,
and other physical processes are lower than their laboratory counterparts, due to the lower pressure and longer interaction timescales in interstellar space
\citep[e.g.][]{BEG_00_13co2}.
Assumptions about particle shapes have not been applied to the laboratory spectra used in this broad analysis, and such corrections would in any case
be inappropriate for the annealed inhomogeneous mixtures used \citep{GWE_99_molcloud}.

\subsection{Spectral Fitting} \label{sec:fitting}
To confirm the validity of using laboratory spectra to model circumstellar material, we estimate the total extinction $A_V$ along the line of sight towards
our YSOs using $A_V$/$\tau_{Si}$ $\sim$ 17 \citep{Rieke_85_extinctionlaw}.  We find $A_V$ $\sim$ 3-34 for our total sample---and $A_V$ $\sim$ 6-34 for those objects
without clear signs of dust emission contaminating the absorption feature---compared with the average 
extinction of $A_V$ $\sim$ 1-5 toward Taurus association stars \citep{Myers_83_densecores}.  
In addition, recent work has shown that in dense regions, this relation with a given $\tau_{Si}$ substantially 
underestimates $A_V$ \citep{Whittet_88_extinctionlaw,Chiar_07_extinctionlaw}, so the total circumstellar extinction is likely to be even higher.

Thus, we assume that the foreground absorption towards the Taurus region is relatively low and that the spectral ice features result from material in the 
circumstellar environment.
In contrast with a least-components type of analysis \citep[e.g.][]{Boogert_08_lowmassices,Pontop_08_co2inlowmass}, 
we determine the ice composition and column densities of the circumstellar ice
by fitting linear combinations of laboratory data to the observed IRS spectra.
With the laboratory data described in Section~\ref{sec:isolate_features_lab}, detailed spectral fitting is performed on the 6.0 and 15.2\mic features using a standard 
$\chi^2$-minimization technique.  
The equation solved is
\begin{equation} \label{eqn:fit}
\tau(\lambda)=\sum{a_i\tau_{lab,i}(\lambda),}
\end{equation}
where the scale factors $a_i$ of each component are the free parameters in the fit.  To compare goodness-of-fit among various combinations, we use $\chi^2_\nu$, the 
reduced $\chi^2$ parameter.  This is defined as $\chi^2_\nu$=${\chi^2}$/(${n-m}$), where $n$ is the number of spectral data points 
and $m$ is the number of free parameters in the fit.
Traditional statistics require $\chi^2_\nu\sim1$ for an acceptable fit, but our values are often significantly larger or smaller.  Artificially small $\chi^2_\nu$
values can be caused by uncertainty overestimation or by poor signal-to-noise.  Large values may be a result of uncertainty underestimation, unknown 
uncertainties in the laboratory spectra, or the existence of components not included in the fits.  
Therefore $\chi^2_\nu$ comparisons may be made to determine
the best laboratory fit to a single spectrum; however, comparisons of goodness-of-fit among different YSO spectra are not necessarily valid.  
Quantities derived using a fitted spectral feature (e.g. processed fractions, column densities) are calculated using an average of all fits for that feature with a
$\chi^2_\nu$ within 20\% of the minimum.

\subsection{Column Densities} \label{sec:column}
To calculate column densities along the lines of sight to the YSOs, we use the relation \citep[e.g.][]{ACA_03_isocam}
\begin{equation} \label{eqn:column}
N(cm^{-2})=\frac{\sum{\tau_\lambda\Delta\lambda}}{\lambda_{peak}^2A},
\end{equation}
where $\Delta\lambda$ is the wavelength bin size, $\lambda_{peak}$ is the wavelength of the feature's peak optical depth, and
$A$ (cm molec$^{-1}$) is the band strength as determined in the laboratory.  

\subsection{SED Spectral Index} \label{sec:alpha}
To provide better resolution in evolutionary status than the simple \citet{Lada_87_starformation} classification, we use the
spectral index $\alpha$ \citep[e.g.][]{KenHart_95_YSOlist}:
\begin{equation} \label{eqn:alpha}
\alpha\equiv\frac{d\log{\lambda F_\lambda}}{d\log{\lambda}}\sim\frac{\log{\lambda_2F_{\lambda_2}}-\log{\lambda_1F_{\lambda_1}}}{\log{\lambda_2}-\log{\lambda_1}},
\end{equation}
where we choose $\lambda_1$ and $\lambda_2$ to lie within well-calibrated and nearly-featureless
parts of the spectrum but as much separated as possible (i.e. $\lambda_1 = 7$ $\mu$m and $\lambda_2 = 30$ $\mu$m; Table~\ref{tab:obj}).
The slope of this section of the spectrum depends upon a variety of factors, including disk radius, total envelope density, outflow cavity
opening angle, and inclination angle.  With the exception of inclination angle, all of these factors depend upon the total dust column density in such a
way that more dust produces a steeper SED slope, and models demonstrate that inclination angle effects are generally less significant than these other factors
\citep{Furlan_08_classImodel}.  Thus small values for the spectral index $\alpha$ indicate a relatively blue or exposed stellar object, but with increasing
$\alpha$ the spectrum becomes redder, indicating the YSO is more heavily embedded in dust.

\section{OPTICAL DEPTH TRENDS} \label{sec:trends}

\begin{figure}
 \begin{center}
 \includegraphics[width=0.45\textwidth]{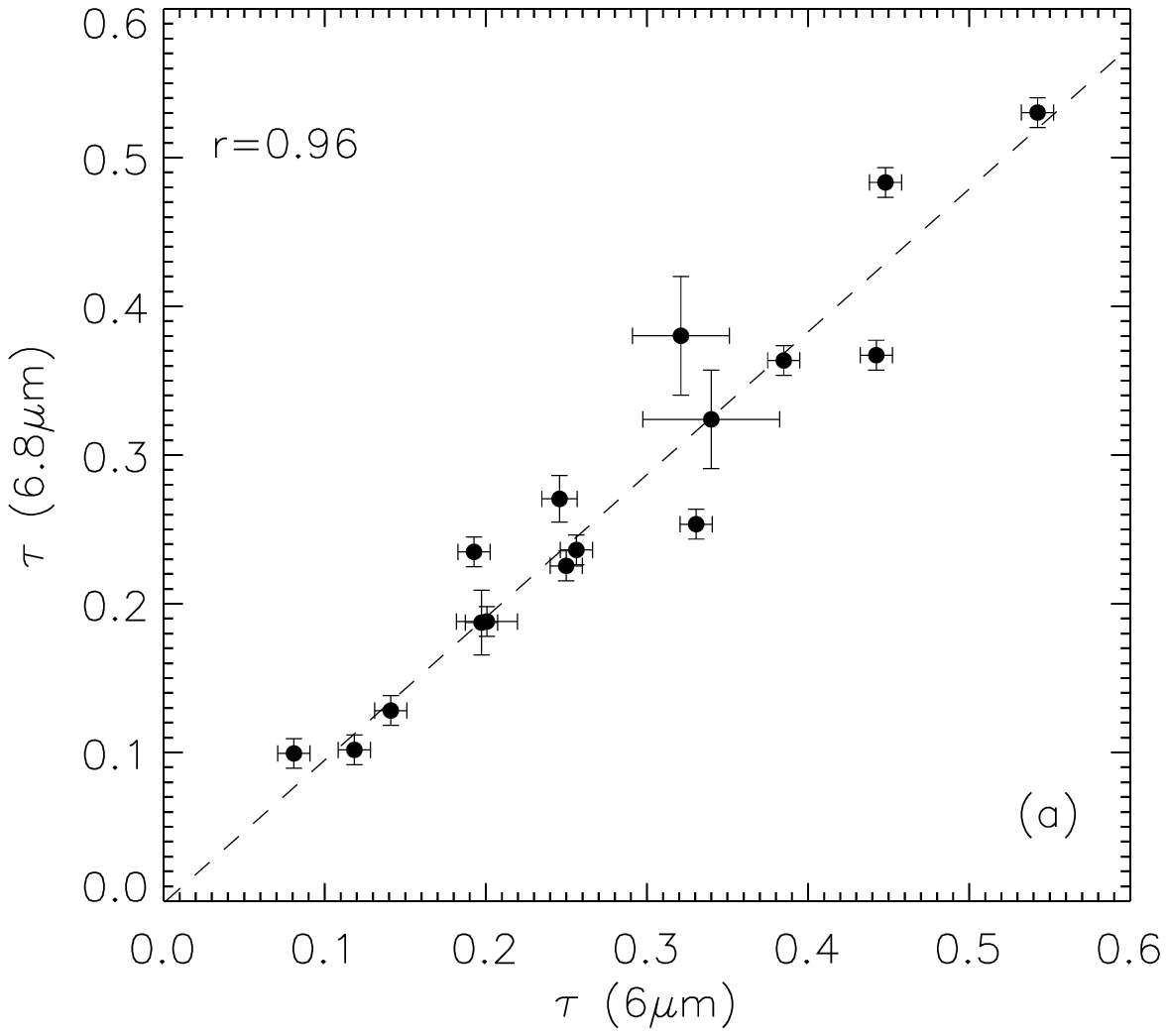}
 \includegraphics[width=0.45\textwidth]{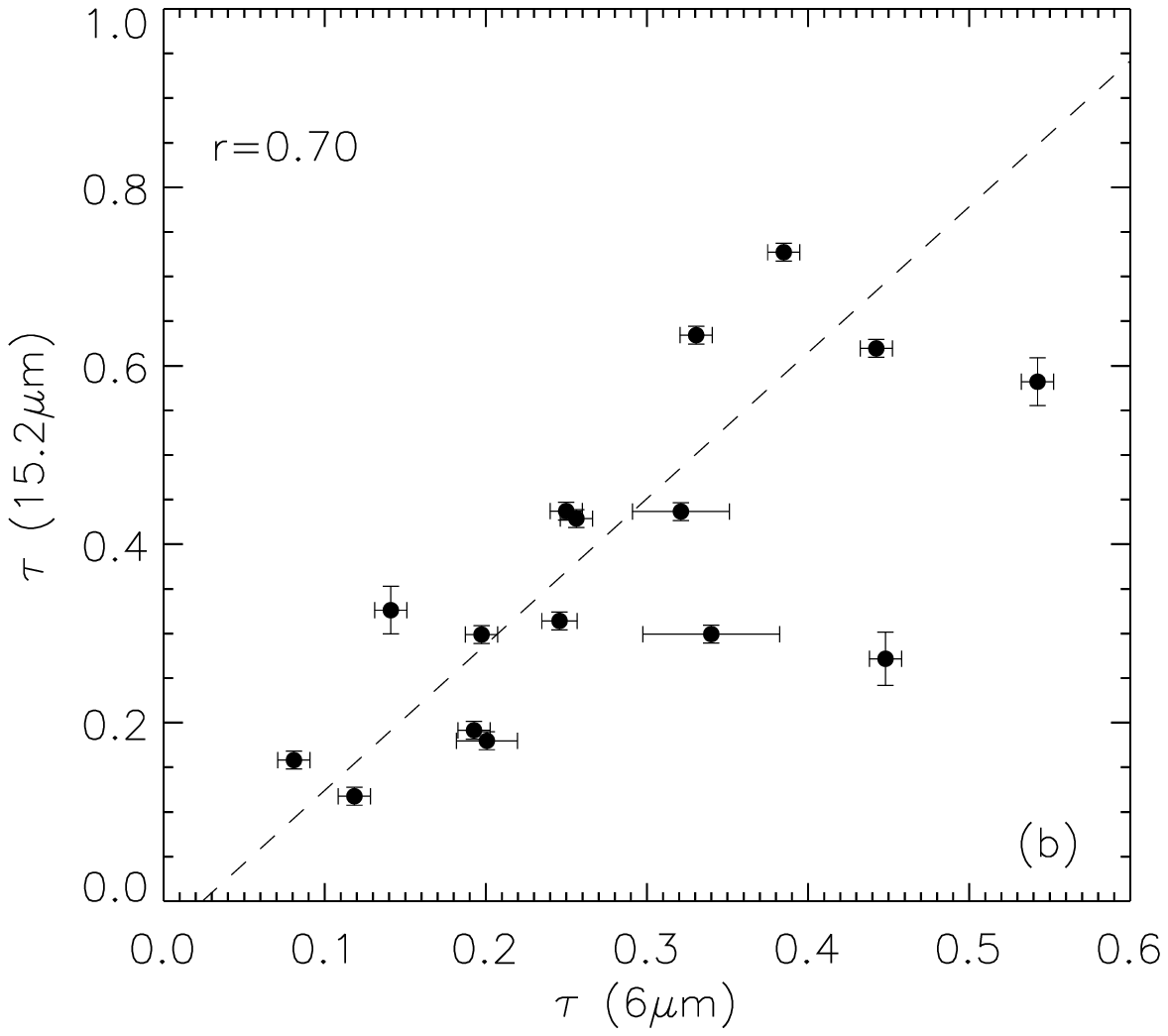} 
 \includegraphics[width=0.45\textwidth]{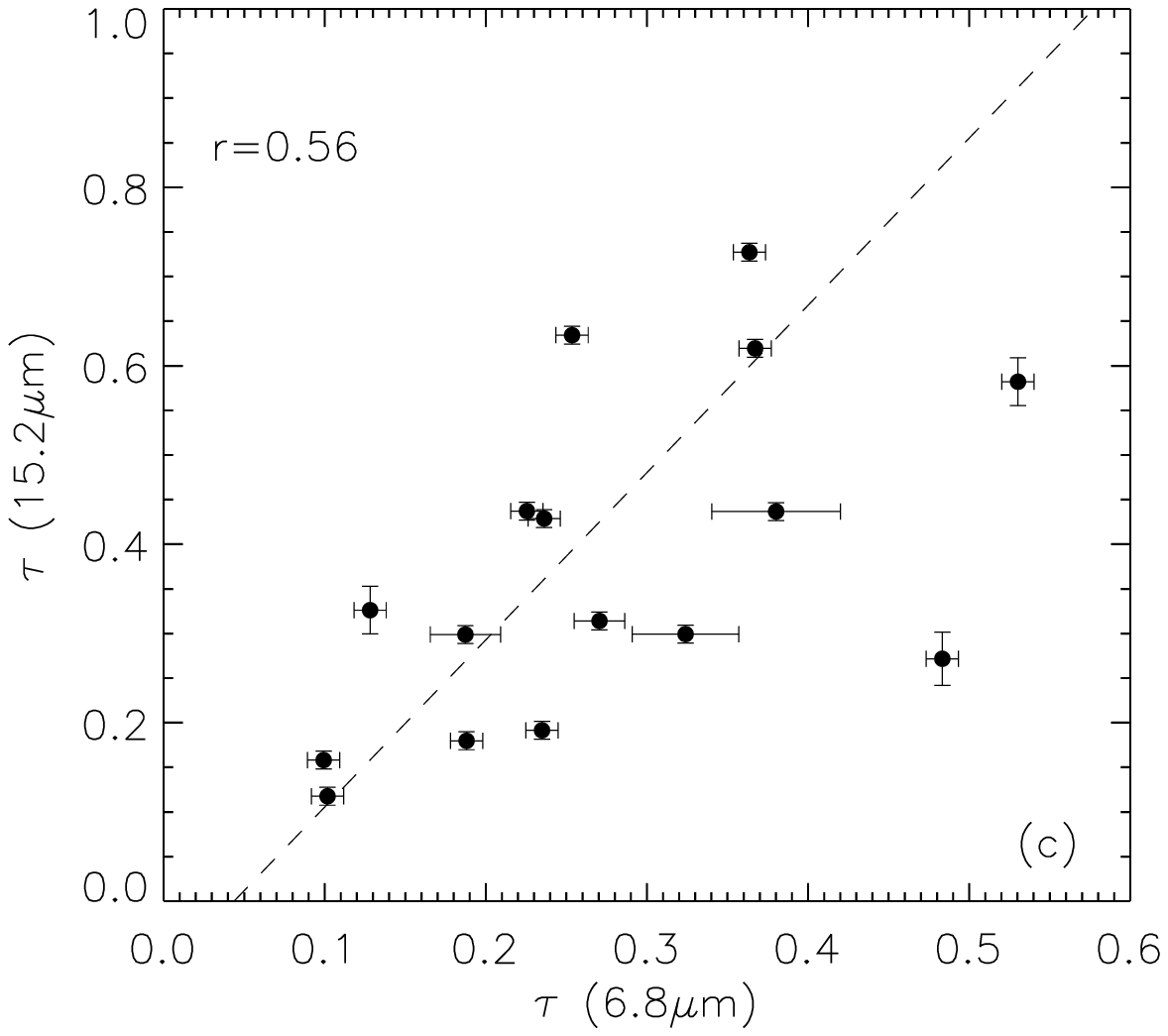} 
 \end{center}
 \caption{Relationships between feature peak optical depths for our low-mass YSOs.  Panel (a)
shows $\tau_{6.8\mu m}$ vs $\tau_{6.0\mu m}$, panel (b) shows $\tau_{15.2\mu m}$ vs $\tau_{6.0\mu m}$, and panel (c) shows $\tau_{15.2\mu m}$ vs $\tau_{6.8\mu m}$.  The
dashed lines indicate the best linear fit.}
 \label{fig:objratios}
\end{figure}

In Figure~\ref{fig:objratios}, we show correlations in the optical depths of the three strongest absorptions
attributed to ices (at 6.0, 6.8, and 15.2 $\mu$m) in our YSO sample.
Clearly, the 6.0 and 6.8 $\mu$m features are 
very closely correlated, with linear Pearson correlation coefficient $r=0.96$ (Figure~\ref{fig:objratios}a).  
The peak ratio is $\tau_{6.8}$/$\tau_{6.0}$ $\sim$ 0.96$\pm$0.03,
and the straight-line fit passes within 0.01 of the origin.
Since the 6.0 $\mu$m band is
primarily attributed to the bending mode of H$_2$O, this correlation indicates
that the carrier(s) of the 6.8 $\mu$m band forms and evolves in a similar, polar environment.   

\begin{figure}
 \includegraphics[]{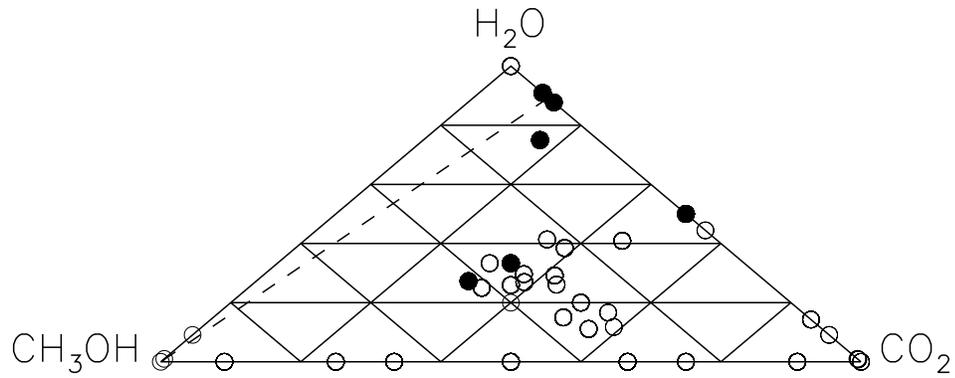}
 \caption{Range of non-irradiated H$_2$O:CH$_3$OH:CO$_2$ laboratory compositions for which there is infrared spectroscopy in the Leiden University database (all circles).
The vertices represent the pure molecular ices, and distance from each 
molecule's vertex indicates decreasing relative abundance of that molecule (solid lines at increments of 20\%).  
The filled circles represent mixtures whose $\tau_{6.0}$/$\tau_{15.2}$ 
ratio falls within the range spanned by the ratios from the YSO spectra (Section~\ref{sec:trends}),
and the dashed line indicates the compositional locus where CO$_2$/H$_2$O=0.12 (Section~\ref{sec:chemdiscuss}).}
 \label{fig:h2oco2_temp}
\end{figure}

There is a looser correlation between the 6.0 $\mu$m and the 15.2\mic optical depths in our 
YSOs ($r=0.7$, Figure~\ref{fig:objratios}b), resulting in a wider ratio range $\tau_{6.0}$/$\tau_{15.2}$ $\sim$ 0.43-1.65.  
When the same ratios are calculated for the laboratory spectra, though the spread is even wider (0.01-3.6), only a small fraction
fall within the range defined by the YSOs (Figure~\ref{fig:h2oco2_temp}, filled circles); the 15.2\mic CO$_2$ feature is generally much stronger than the 
6.0\mic water ice feature, independent of temperature. Mixtures dominated by H$_2$O provide most of the closest matches.
In addition, there are two lab mixtures of nearly 1:1:1 H$_2$O:CH$_3$OH:CO$_2$ abundances (center of Figure~\ref{fig:h2oco2_temp}) which have 
$\tau_{6.0}$/$\tau_{15.2}$ ratios comparable to the YSO ones but only after they have been heated to high temperatures (T $\gtrsim$ 125 K),
nearing the ice sublimation points.
The Leiden database has a scarcity of water-rich, CO$_2$-poor ($\lesssim$20\%) H$_2$O:CH$_3$OH:CO$_2$ mixtures in the temperature range
10-180 K, though based on this comparison, these compositions appear to be astronomically relevant.  Because of the undersampling of compositions in this range, we
are not able to constrain the ice composition well from the $\tau_{6.0}$/$\tau_{15.2}$ ratio alone.

Despite being closely correlated with $\tau_{6.0}$, the 6.8\mic peak optical depth is even less correlated with the 15.2\mic depths 
($r=0.56$, Figure~\ref{fig:objratios}c).  This indicates that much of the inherent scatter in the 
$\tau_{6.8}$/$\tau_{6.0}$ ratio is due to variations in the 6.8\mic feature among different lines of sight.

\section{SPECTRAL FEATURE ANALYSES} \label{sec:features}

In order to set more accurate constraints on the ice properties than feature-strength analysis provides, we perform spectral fitting and detailed analysis
of the most prominent features.

\subsection{The 6.0 $\mu$m Feature} \label{sec:6um_feature}

\begin{figure}
 \begin{center}
 \includegraphics[width=0.8\textwidth]{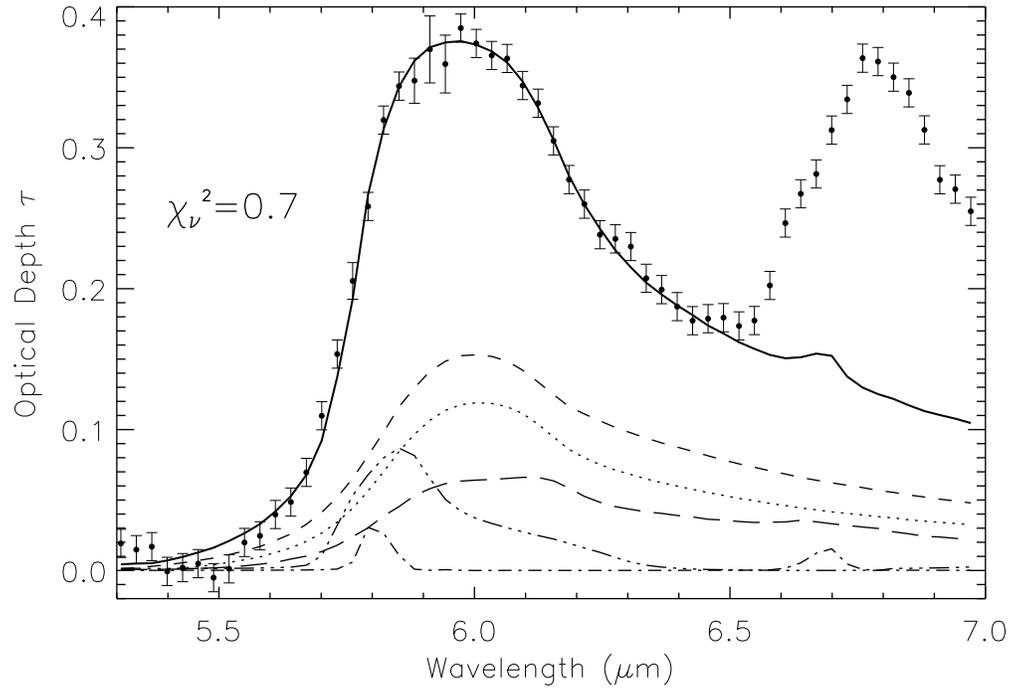}
 \end{center}
 \caption{The 6\mic feature of IRAS 04016+2610, fit by a combination of pure H$_2$O at 10 and 50 K (dots and short dashes), pure H$_2$CO at 10 K (dot-dash), 
pure HCOOH at 10 K (dash-triple dots), and H$_2$O:NH$_3$ 10:2 at 10 K (long dashes).  The solid line is the sum.  The 6.8\mic feature is deliberately excluded from the fit.}
 \label{fig:6umfit}
\end{figure}

Each YSO spectrum is fitted over the wavelength range 5.4-6.3\mic using the method described in 
Section~\ref{sec:fitting} and the component set listed in Table~\ref{tab:6um_components}.  An example result is shown in Figure~\ref{fig:6umfit}, and the total
sample residuals, after subtraction of the 6\mic components, are shown in Figure~\ref{fig:6.8um_spectra}.  
Due to overlap with the unidentified 6.8\mic feature, the range beyond 6.3\mic was excluded in order to avoid imposing artificial structure on any fit residuals.  
These fits focus on identifying the constituents of the 6.0\mic feature---H$_2$O, H$_2$CO, HCOOH, and NH$_3$, as described in the following subsections.
This suite of species is similar to that employed by \citet{Boogert_08_lowmassices} to explain their extracted component set C1-5.

Either noise or saturation of the absorption bands in the background star/Class 0 sample prevents detailed spectral fitting of the 6\mic features, 
so we scaled a pure H$_2$O (10 K) spectrum to match the peak optical depth at 6.0 $\mu$m in order to calculate $N$(H$_2$O) (Table~\ref{tab:molcloud_abun}).
The shape of the 6\mic W3 spectrum is too irregular to be reasonably fit with a solid H$_2$O lab spectrum in this manner.  
 
\subsubsection{H$_2$O} \label{sec:h2o}

Water is the least volatile and most abundant of the ices typically found along protostellar lines of sight. The mid-infrared spectrum offers three strong features of 
water ice: narrow ones at 3.1 and 6.0 $\micron$, and a broad ($\lambda$ $\sim$ 11-20 $\mu$m) feature peaking at 13 $\micron$. 
To turn these profiles into water ice column 
density is somewhat problematic in each case. The 3.1 $\micron$ ice feature is generally regarded as the cleanest measure of H$_2$O ice column density 
\citep[e.g.][]{Tielens_84_5-8micspectra}. 
However, inconvenient gaps in the spectrum transmitted by Earth's atmosphere make it difficult to determine the continuum on the short wavelength side of this 
feature in ground-based observations, which would be the source of most of the comparisons we could make to our data. In addition, rather significant extinction and 
scattering near 3 $\micron$ are inevitably associated with protostars and can affect the profile of the water ice feature. 
Futhermore, this may lead to the bulk of the 3\mic background continuum taking a significantly different path through the protostellar envelope 
than that which gives rise to the longer-wavelength features. The latter 
effect would be rather more serious in systems viewed closer to edge-on, such as IRAS 04169+2702 and DG Tau B. 

Extinction at 6 $\micron$ is similar to that at 13 $\micron$, and both are significantly smaller than that at 3 $\micron$ \citep[extinction in magnitudes smaller by 
factors of $\sim$ 2-3;][]{Mathis_90_extlaw,Indeb_05_extlaw}. 
Thus the optical paths sampled at 6 and 13 $\micron$, and the ice column densities derived, should be similar and may be 
more characteristic of the line of sight to the embedded protostar than is the case at 3 $\micron$. 
In observed spectra, the 6 $\micron$ feature is the easiest of the three to analyze, as there are fewer problems in determination of the continuum. However, the 
depth and profile of the 6 $\micron$ feature are possibly influenced by absorption from species other than water ice, as shown, 
for example, by \citet{Boogert_08_lowmassices}. The 
13 $\micron$ water ice libration feature is broad and overlaps many other absorptions, including the loosely-constrained 9.7\mic silicate feature, 
so its optical depth and feature shape are hard to measure precisely.

Taking account of all of these difficulties, we choose to measure water ice column density with the 6 $\micron$ feature, and check against the possibility of substantial 
``contamination'' from other species using the 13\mic feature.  
To calculate the H$_2$O column density, we use the 6.0 $\micron$ feature fit results along with 
Equation 2 (summed over $\lambda$=5.5-7 $\mu$m), with a band strength of $A_{H_2O}$=1.2$\times$10$^{17}$cm molec$^{-1}$ for the
T = 50 K water ice components and $A_{H_2O}$=1.3$\times$10$^{17}$cm molec$^{-1}$ for the T = 120 K one. These values pertain to pure ${\rm H_2 O}$ ice, though they are 
also suitable for apolar dilutant concentrations of $\lesssim$30\% \citep{GSG_95_labsim}. We note that the derived column densities $N({\rm H_2 O})$ (Table 5) may in 
some cases represent upper limits, as the overall composition of the 6 $\micron$ feature is still a source of discussion \citep[e.g.][]{GWB_04_ice,Boogert_08_lowmassices}. 
However, as we will see below, the consistency of these results with those from 13 $\micron$ features in our spectra indicates that the 
contribution of unaccounted-for species at 6 $\micron$ is rather small compared to that of water ice. 
In general, our values of water ice column density are somewhat larger than those derived for the same objects from 
the 3.1 $\micron$ feature (e.g. Boogert et al. 2008), and tend to depart more the closer the object is to an edge-on view, 
so the differences may in part be due to the expected differences between the light paths viewed at the two wavelengths.

\begin{figure}
 \begin{center}
 \includegraphics[width=0.7\textwidth]{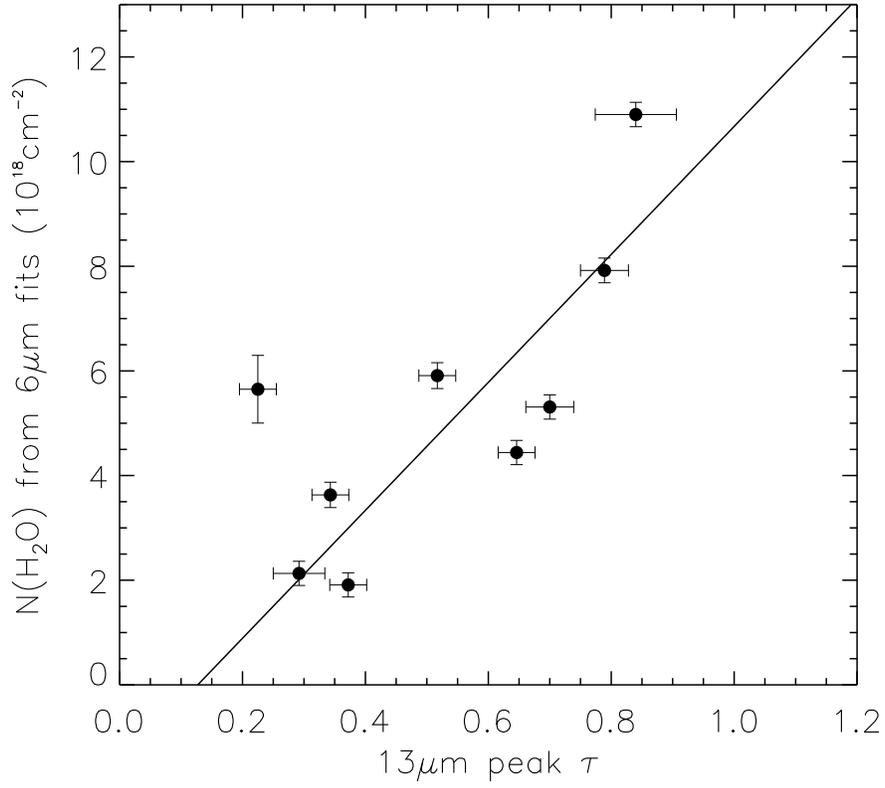}
 \end{center}
 \caption{Peak optical depth in the water ice libration mode ($\lambda$ $\sim$ 13 $\micron$), plotted against the water ice column density $N$(H$_2$O) 
derived from the 6 $\micron$ water ice feature as described in Section~\ref{sec:h2o}.  Points and uncertainties are results for the nine Taurus objects 
in which the 9.7\mic silicate features are strongly in absorption.}
 \label{fig:libh2o}
\end{figure}

To look for signs of additional ice components with features near 6 $\micron$ that are not resolved spectrally from the water-ice feature, we have measured the 11-20 
$\micron$ solid H$_2$O absorption in the nine targets for which the 10 $\micron$ silicate feature is most deeply in absorption. 
To do so, we followed a procedure similar to 
that by \citet{Boogert_08_lowmassices}. From a low-order polynomial fit to the continuum, constrained to follow the spectrum at 5.3-5.6 and 30-35 $\micron$, we determined 
the total optical depth in the 8-30 $\micron$ range. To this we ``fit'' the profile of interstellar silicate absorption, exemplified by the galactic center source GCS 3 
\citep{Kemper_04_silicates} and constrained to match the observed spectrum at 9.9-10.1 $\micron$ and 23-30 $\micron$. 
Since this interstellar silicate profile may not be fully applicable to circumstellar silicates, this step introduces additional uncertainty.
Subtraction of the resulting profile from the optical 
depth spectrum reveals the water ice libration absorption, peaking at $\sim $13 $\micron$, and interrupted at 15.2 $\micron$ by the $\rm{CO}_2$ ice feature. We measured 
from such results the peak optical depth in this water ice feature around 13 $\micron$. By small variation of the wavelengths at which the continuum and silicate profile 
were constrained to follow the spectrum, we estimated the uncertainty in the peak optical depth. Because the shape of this very broad water feature is more sensitive to 
both the continuum and silicate profile fits than is the peak optical depth, the corresponding water 
ice column density is considerably more uncertain.  In Figure~\ref{fig:libh2o}, 
the values for peak optical depth are plotted against the column density $N$(H$_2$O) derived from the 6 $
\micron$ feature as described above. Evident in Figure~\ref{fig:libh2o} is a good correlation between the 13 and 6 $\micron$ water signatures, with Pearson's $r~=~0.74$.
That the correlation is tight indicates that, if an additional ice component is significant at 6 $\micron$, 
its abundance relative to water ice must be essentially constant.

\subsubsection{H$_2$CO and HCOOH} \label{sec:h2co_hcooh}
Many of the objects in our sample have additional absorptions centered around $\sim$5.8\mic that can be well-fit by a combination of 
H$_2$CO (formaldehyde) and HCOOH (formic acid) ices.  Solid H$_2$CO is observed in many protostellar systems 
\citep[e.g.][]{KTB_01_5-8umregion,GWB_04_ice} and plays an important role in the production of complex organic molecules \citep{Schutte_93_h2co}.
H$_2$CO is difficult to detect independently in YSO Spitzer spectra; its
3.47\mic C-H stretch feature is beyond the wavelength range of the IRS, and its weaker 6.68\mic C=O stretch feature is blended with the 6.8\mic absorption,
itself not well understood.  The H$_2$CO column densities (Table~\ref{tab:abundances}) are derived 
using the observed integrated optical depth of the 5.8\mic feature (as determined from
the spectral fits) and a band strength of $A_{H_2CO}$=9.6$\times$10$^{-18}$cm molec$^{-1}$ \citep{Schutte_93_h2co}.

Solid and gas-phase HCOOH has been detected in both low- and high-mass star-forming regions 
\citep[e.g.][]{Cazaux_03_hcoohdetect,Schutte_99_7.2/7.4mic,Bottinelli_04_hcoohdetect,Remijan_04_hcoohdetect,KTB_01_5-8umregion}.
Within the IRS wavelength range, there are solid-state HCOOH absorptions at 5.8 and 7.2 $\mu m$, and several in
the 8-11\mic range \citep{Bisschop_07_hcooh}.  The latter ones are blended with the 9.7\mic silicate complex, but the 7.2\mic band,
though intrinsically $\sim$4 times weaker than the 5.8\mic resonance, is detectable and indeed is 
tentatively identified in 6 of our 16 objects (Section~\ref{sec:7-8mic}).  We have derived HCOOH column densities (Table~\ref{tab:abundances}) by integrating the 
observed optical depth of the 5.8\mic feature (as determined from the spectral fits) and using a band strength of $A_{HCOOH}$=6.7$\times$10$^{-17}$cm molec$^{-1}$ 
\citep{Marechal_87_hcooh}.

\subsubsection{NH$_3$} \label{sec:nh3}
Solid NH$_3$ (ammonia) has mid-infrared features at 6.15\mic and 9.3 $\mu$m \citep[e.g.][]{SA_93_ch3ohnh3}, 
blended with the 6\mic water ice and 9.7\mic silicate features, respectively.  The effect on the 6\mic H$_2$O feature is negligible until the NH$_3$ concentration 
reaches $\sim$9\% \citep{KTB_01_5-8umregion}.  At higher concentrations, an excess appears around 6.15\mic on the red side of the 6\mic band in laboratory spectra,
which fits well the broad 6\mic feature observed in many of our YSO spectra.  
Analyses of the 9\mic inversion mode resonance by \citet{Lacy_98_frozennh3} and
\citet{GWS_00_W33A} indicate that circumprotostellar NH$_3$ exists primarily in an H$_2$O-dominated matrix.
Therefore, an H$_2$O:NH$_3$ (1:0.2) mixture, rather than pure NH$_3$,
is used in the spectral fits.  Because no satisfactory band strength measurements for NH$_3$ in such a mixture exist,  
the upper limit NH$_3$ column densities (Table~\ref{tab:abundances}) are estimated by using the band strength 
of pure water ice for the H$_2$O:NH$_3$ mixture and extrapolating
the column density of NH$_3$ based on its fractional abundance in the final fit.

\subsubsection{PAHs} \label{sec:remnants}

After contributions due to H$_2$O, H$_2$CO, HCOOH, and NH$_3$ are subtracted, small residuals in the 6.1-6.4\mic range, varying in strength, shape, and peak position,
are observed in 11 of the sample's 16 YSOs (Figure \ref{fig:6.8um_spectra}).  PAH absorption has been suggested as a candidate for the 6.2-6.3\mic 
feature observed in many massive protostars \citep{Schutte_96_6/6.8inngc7538}.  Using the average interstellar PAH spectrum
compiled by \citet{Hony_01_PAHspectrum} as a reference, we examined the peak position and widths of the YSO residuals.  Of the objects within
our sample with excess absorptions peaking between 6.2 and 6.3 $\mu$m, only three (IRAS 04361+2540, DG Tau B, and HL Tau) have an excess with a width similar to 
that seen in the PAH spectrum, and none of these three match the PAH feature in shape and peak position.  
Though the PAH relative band strengths often depend upon molecular environment and ionization level, the broad 7.5-9\mic complex and the 6.2\mic feature are 
consistently observed at comparable strength in astronomical environments \citep{Hony_01_PAHspectrum}, yet we find no correlation
in the YSO spectra between the 6.2 and 7.7\mic optical depths---the ratios are scattered between 0 and 7.5.
Moreover, the prominent but narrow PAH peak at 11.3\mic \citep[consistently $\sim$1.4$\times$ as strong as the 6.2\mic peak;][]{Chan_00_PAHratios,Hony_01_PAHspectrum} 
and the lesser ones at 12.7, 13.5, and 14.3\mic are not observed in the YSO data.  Thus in agreement with \citet{KTB_01_5-8umregion}, 
we find it unlikely that PAHs contribute significantly around 6.2-6.3 $\mu$m.

\subsection{The 6.8 $\mu$m Feature} \label{sec:6.8um_feature}

Though the 6.8\mic interstellar absorption band was discovered nearly 30 years ago \citep{Puetter_79_6.8umdiscovery}, 
its identification remains ambiguous and inconclusive.  
Species ranging from hydrocarbons to alcohols to carbon dust have been proposed to contribute to the feature in a variety of sources
\citep[see][]{KTB_01_5-8umregion}.
Here we focus on CH$_3$OH and NH$_4^+$---the candidate carriers most commonly accepted as explaining at
least part of the absorption \citep[e.g.][]{BE_04_ISI}.

\subsubsection{CH$_3$OH}

The position of the 6.8\mic feature coincides with the C-H deformation mode of alcohols, and of these, methanol is the most likely to be found in interstellar abundance.
Theoretical models of ice processes predict the creation of solid methanol through the sequential hydrogenation of CO ice which has been accreted 
from the gas phase onto dust grains \citep{Tielens_97_methproduction}.  
The presence of methanol ice in interstellar and protostellar environments is observationally supported by detection of
its own resonances---e.g. the C-H stretching mode at
3.54\mic \citep{Grim_91_methdetect} and the C-O stretch at 9.8\mic \citep{GWS_00_W33A}---as well as by signatures of its interactions with other molecules
\citep[especially CO$_2$; e.g.][]{EDD_98_segregation}. 

Our sample spectra do not contain the NIR wavelengths of CH$_3$OH's multiple stretching modes, and the 9.8\mic resonance is blended with the silicate
absorption.  However, constraints on the CH$_3$OH abundance can be derived using the 15.2\mic CO$_2$ band, as the interaction of CH$_3$OH with CO$_2$ changes
the feature shape near 15.4\mic (Sections~\ref{sec:15um_feature} and \ref{sec:chemdiscuss}).  The relative strength of the 15.4\mic shoulder implies
that at most 50\% of the 6.8\mic feature is due to CH$_3$OH in a CO$_2$ matrix, although CH$_3$OH unmixed with CO$_2$ could be present
to provide additional absorption at 6.8 $\mu$m. 
Since CH$_3$OH is thought to form in a
hydrogenated environment, the assignment of at least part of the 6.8\mic absorption to this molecule is then supported by 
the close correlation of the 6.0 and 6.8\mic optical depths (Section~\ref{sec:trends}).

\subsubsection{NH$_4^+$}
The NH$_4^+$ ion was proposed as the carrier of the 6.8\mic band by \citet{Grim_89_nh4plus}, and since then, this proposal has achieved varying levels of success
\citep[e.g][]{Schutte_96_6/6.8inngc7538,KTB_01_5-8umregion,Schutte_03_nh4plus_6.8um}.  The ion can be produced in
the laboratory by the acid-base reaction HNCO + NH$_3$ $\rightarrow$ OCN$^-$+ NH$_4^+$ \citep{Keane_97_makenh4plus} or
in UV-irradiated H$_2$O:CO:NH$_3$ mixtures (as in Figure~\ref{fig:6.8um_spectra}).  

NH$_4^+$ has expected counter-ion features at 4.62\mic (OCN$^-$) and 6.3 and 7.41\mic (HCOO$^-$, produced by photolysis).  The 4.62\mic 
feature does not fall within the IRS wavelength coverage, and the 7.41\mic signature is too weak to be conclusively identified in our data.
However, some of our sample
objects do have a small excess absorption near 6.3 $\mu$m (Section~\ref{sec:remnants}), which is unlikely to be attributable to PAHs.  
It may instead be due to HCOO$^-$, indicating photolysis and 
subsequent warming of a polar ice containing NH$_3$.  We use the ratios of the laboratory 6.3\mic (HCOO$^-$) and 6.8\mic (NH$_4^+$) peaks 
($\tau_{6.8}/\tau_{6.3}\sim1.7$ after the H$_2$O feature is removed), along with the equivalent widths of the YSO and 
laboratory features, to estimate the NH$_4^+$ contribution to the 6.8\mic feature.  Using this method, we find that, within our sample, 6-58\% 
of the 6.8\mic feature can be attributed to NH$_4^+$ (HL Tau is an outlier whose particularly large 6.3\mic excess indicates $\lesssim$91\% of its
6.8\mic feature may be due to NH$_4^+$ using this method).

\subsubsection{Fits and Summary} \label{sec:6.8micfits}

\begin{figure}
 \begin{center}
 \includegraphics[width=0.75\textwidth]{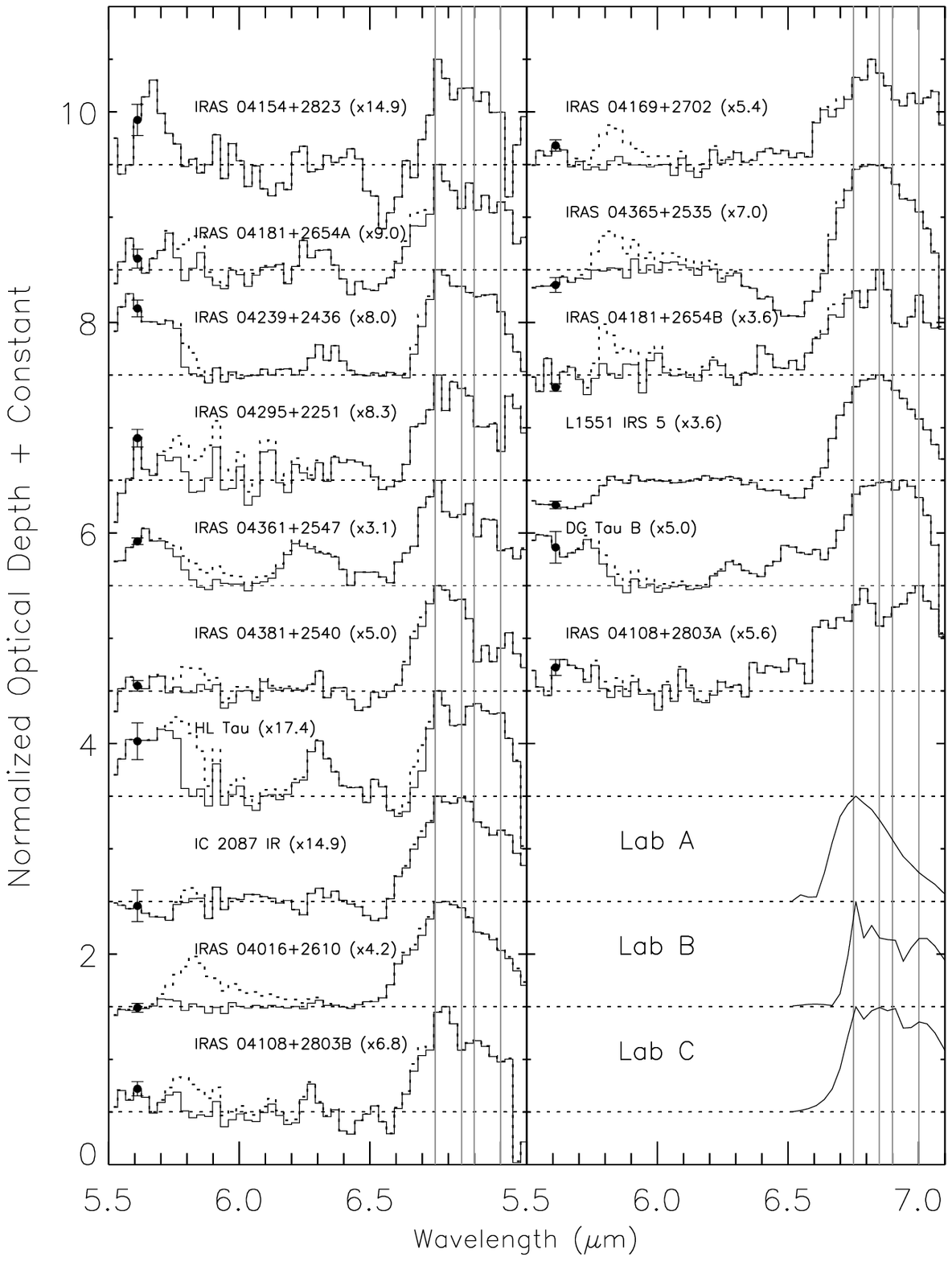}
 \end{center}
 \caption{The sample's 6-7\mic profiles (after subtraction of the H$_2$O contributions (dashed) and all contributions (solid) at 6 $\mu$m) 
normalized to 1 and shifted vertically, sorted by 6.8\mic peak position (the point near 5.6\mic indicates the approximate optical depth uncertainty).
The 5.6-5.7\mic residuals are almost entirely byproducts of continuum mismatches.  
Also shown for comparison are normalized laboratory spectra---{\it Lab A}: H$_2$O:CO:NH$_3$ 10:2.5:2 at 165 K (irradiated to create NH$_4^+$), 
{\it Lab B}: H$_2$O:CH$_3$OH:CO$_2$ 9:1:2 at 10 K, {\it Lab C}: H$_2$O:CH$_3$OH 1:100 at 10 K.}
 \label{fig:6.8um_spectra}
\end{figure}

Figure~\ref{fig:6.8um_spectra} shows a 
comparison of the 6.8\mic features, after subtraction of the 6\mic contributors, with a variety of 6.8\mic candidate laboratory spectra,
all rebinned to the resolution of the observations.  The bottom two lab spectra (Lab B and C) have CH$_3$OH/H$_2$O abundance ratios of 100 and 0.1,
and the gray vertical lines indicate the peak positions of these mixtures.  The relative amount of methanol present affects the peak substructure.
While in some cases the YSO peak positions and relative strengths closely resemble the laboratory data (e.g. IC 2087 IR and IRAS 04239+2436),
in others there is no correlation at all (e.g. IRAS 04169+2702).
The abundance of CH$_3$OH (relative to H$_2$O) has been found to vary by almost an order of magnitude among different protostellar lines
of sight \citep[e.g][]{Pontopp_03_methdetect,GWB_04_ice}, which may partly explain the observed variety of peak substructures.

The top laboratory spectrum in Figure~\ref{fig:6.8um_spectra} (Lab A) is a H$_2$O:CO:NH$_3$ deposit, 
irradiated to create NH$_4^+$ and subsequently warmed from 10 to 165 K.
The irradiated spectrum at 10 K peaks 0.1\mic bluewards of our sample's bluest 
feature; in order to align with any of the observed
profiles, the irradiated mixture must be heated to near the H$_2$O sublimation temperature,
as the sublimation appears to cause the redward peak shift \citep{Schutte_96_6/6.8inngc7538}.  Moreover, the width of
the NH$_4^+$ feature is too narrow ($\sim$0.25 $\mu$m) to account for the entire 6.8\mic feature widths observed towards each YSO (average $\sim$0.4 $\mu$m).  Even if
multiple NH$_4^+$ laboratory spectra at various temperatures are combined in order to reproduce the observed feature width, 
the resultant peak position remains too blue to match the YSO data.

Noting that many of the YSO 6.8\mic features share characteristics with the profiles of both CH$_3$OH (triple peaks, 7\mic shoulder) and NH$_4^+$ (asymmetric peak near
6.75 $\mu$m), we attempted fitting them with a combination of the laboratory components.
While no remarkably good fits were achieved, there emerged a general trend that equal contributions from NH$_4^+$ and from CH$_3$OH in a water-dominated matrix, 
with varying amounts of nearly-pure CH$_3$OH intermixed, came closest to reproducing many of the 
YSO feature width-to-depth ratios (i.e. mixtures of irradiated NH$_3$, CH$_3$OH/H$_2$O$\sim$0.1, and CH$_3$OH/H$_2$O$\sim$100 
in the approximate ratios 1:1:0 to 1:1:1).  
However, the overall 6.8\mic astronomical feature shapes could not be be well-matched in terms of either peak alignment or subpeak relative strengths.

Based on this work, we conclude that we cannot rule out NH$_4^+$ and CH$_3$OH as contributors each of at most $\sim$50\%
to the 6.8\mic band, but we emphasize that even the 50/50 combinations do not reproduce the overall feature shapes.  
Studies of this feature performed with higher-resolution ISO-SWS observations 
\citep[e.g.][]{KTB_01_5-8umregion,Schutte_96_6/6.8inngc7538} find a similar variety of structure and peak positions, confirming that it may be difficult or physically
impossible to adequately explain the feature using the same set of components for all sight-lines.

\subsection{The 7-8\mic Range} \label{sec:7-8mic}

Because of the uncertainties due to the low resolution and low signal-to-noise of the 7-8\mic optical depth spectrum, 
as well as the inherent narrowness of the 7-8\mic features, 
we did not attempt to fit the shape of the spectral features with laboratory profiles but rather focused on detection of the 7.25\mic HCOOH, $\sim$7.6\mic SO$_2$, 
and 7.7\mic CH$_4$ features.

\subsubsection{HCOOH}

The 7.25\mic HCOOH band is intrinsically $\sim$4 times weaker than the 5.85\mic one \citep{Bisschop_07_hcooh}.  
Based on the amount of HCOOH in the fitted spectra in Section~\ref{sec:6um_feature}, the expected 7.25\mic optical depths in the YSO spectra
are $\lesssim$0.02.  Indeed, we do observe small features at 7.25\mic in 6 of our 16 objects.  It should be noted, however, that the strength of these do not
correlate with the HCOOH column densities derived from the spectral fits and the stronger 5.85\mic feature ($r=0.06$);
there are also other species with weak
signatures very close to this wavelength \citep[e.g. HCOO$^-$ and HCONH$_2$;][]{Schutte_99_7.2/7.4mic}.  As the optical depth profiles here are neither strong
nor resolved enough to conclusively distinguish among these species' signatures, these HCOOH detections must remain tentative.

\subsubsection{CH$_4$ and SO$_2$} \label{sec:ch4so2}

In 13 of our 16 objects, we observe a feature near 7.7 $\mu$m, where the ($\nu_4$) C-H deformation mode of CH$_4$ is expected.  Laboratory studies have shown that this
resonance's position and width are strongly dependent on the molecular environment, becoming blue-shifted (as far as 7.64 $\mu$m) and 
broadened when mixed with a variety of other ices, including H$_2$O, CO, CH$_3$OH, and CO$_2$ \citep{Boogert_97_labch4}.
In addition, there is the possibility of blending with the nearby SO$_2$ feature, whose 7.58\mic S-O stretching band 
has been shown to vary its peak position from 7.45-7.62 $\mu$m, 
depending on the temperature and the molecular environment \citep{Boogert_97_labch4}.  

In order to disentangle absorption due to SO$_2$ and CH$_4$, we fit each YSO's 7.45-8\mic wavelength range
with two Gaussian curves, allowing the peak of one (representing SO$_2$) to shift between 7.45-7.62\mic and constraining the other peak (CH$_4$) to 7.62\mic or larger.
A Gaussian approximation was chosen because both the SO$_2$ and CH$_4$ profiles are too narrow, compared with the IRS resolution, to
evaluate any substructure.
In many cases, a single Gaussain was sufficient, indicating a single, distinct feature of either SO$_2$ or CH$_4$, depending on peak position.
Of our 16 YSOs, 7 show only a 7.7\mic feature with little or no asymmetrical excess, 3 only have a distinct feature peaking 
within the SO$_2$ range, 4 show both distinct features, and 2 show a broad absorption with additional peaks
within the SO$_2$ range (i.e. fit by overlapping Gaussians).  Using the range of laboratory feature FWHMs
measured by \citet{Boogert_97_labch4}, we find that in 4 of the 9 objects with either distinct
features or a Gaussian fit in the 7.45-7.62\mic range, the feature width is too narrow ($\Delta\lambda\lesssim$0.06 $\mu$m) to be attributed to solid SO$_2$.  
These may be due to random noise creating an apparent peak or to CH$_4$ gas lines found between $\sim$7.4-8 $\mu$m.
SO$_2$ also has gas lines in this region, but these are very weak \citep{Boogert_97_labch4}.  For the 5 remaining features, we use 
Equation~\ref{sec:column} with a band strength $A_{SO_2}$=3.4$\times$10$^{-17}$cm molec$^{-1}$ \citep{Khanna_88_SO2bandstrength} to determine the SO$_2$ column
density; for the other objects, an upper limit to $N$(SO$_2$) is determined (Table~\ref{tab:abundances}).

Of the 13 objects with a distinct 7.7\mic CH$_4$ feature or Gaussian fit, 7 have FWHMs comparable to those observed in laboratory mixtures \citep{Boogert_97_labch4}, 
particularly in those mixtures where CH$_4$ and either H$_2$O, CH$_3$OH, or NH$_3$ exist in roughly equal abundance.  The remaining 6 objects have wider features,
which may be explained by variations in temperature or ice mantle composition along the line of sight.  To determine the column density of CH$_4$ towards each YSO, 
we use Equation~\ref{sec:column} with a band strength $A_{CH_4}$=7.3$\times$10$^{-18}$cm molec$^{-1}$ \citep[pure CH$_4$; discussion in][]{Boogert_97_labch4} and
integrate the optical depth between 7.6-7.9 $\mu$m.  Upper limit estimates for the 3 non-detections are provided by 
the noise in the spectrum (Table~\ref{tab:abundances}).

\subsection{The 15.2 $\mu$m Feature} \label{sec:15um_feature}

Double-peaked structure in the 15.2\mic feature, often observed in systems with an exposed source of heating and radiation 
(such as Class I/II objects), is associated with the dual peaks of the degenerate $\nu_2$ bending mode in pure CO$_2$ ice.  
This structure becomes less pronounced or disappears altogether when the apolar CO$_2$ molecules are embedded in an H$_2$O-rich polar matrix 
\citep[e.g.][]{EKS_99_laboratory}.  Thus the presence of a double-peaked feature indicates that thermal processing has occured, segregating the CO$_2$ from the polar
matrix.  The broad shoulder
on the red side of the absorption, centered around 15.4 $\mu$m, has been attributed to CO$_2$ interactions with other molecules, particularly CH$_3$OH,
and the strength and shape of the shoulder depend on both temperature and relative abundances \citep[e.g][]{DDd_99_co2ch3oh}.
Theoretical studies suggest that solid CO$_2$ is formed by oxidation of CO ice
\citep[e.g.][]{Tielens_82_compositionmodels,Ruffle_01_co2formation}, so the co-existence and interaction with solid CH$_3$OH 
\citep[also formed from CO;][]{Tielens_97_methproduction} is not unexpected.

In the fits, made for the high-resolution spectra\footnote{The low-resolution spectra are insufficient to resolve any structure which might be present.} 
between 14.7 and 16 $\mu$m, we include cold (10-30 K) deposits of H$_2$O:CO$_2$, pure CO$_2$, and an irradiated deposit of CO:H$_2$O, as well as
warmer ($\gtrsim$100 K) mixtures of H$_2$O:CH$_3$OH:CO$_2$ (Table~\ref{tab:15um_components}).  
The cold mixtures containing H$_2$O are used to estimate the contribution from 
nonprocessed material along the line of sight.  The warmer H$_2$O:CH$_3$OH:CO$_2$ spectra represent ices which have been warmed and have undergone 
the chemical and physical changes, including segregation of the CO$_2$, resulting from that processing.  
The pure CO$_2$ component, though measured at 10 K, does not greatly differ in shape until sublimation at $\sim$80 K 
and is here used as a measure of segregated, highly-processed ice.

Our fitting components differ from the recent work by \citet{Pontop_08_co2inlowmass}, 
which is focused almost exclusively on detailed methanol-free CO:CO$_2$:H$_2$O mixtures. Certainly these combinations can produce good fits to the CO$_2$ ice feature.
However, in the following we will adhere to 
the conventional approach involving a warm CH$_3$OH ice component, since we find that in most cases this produces better fits to our spectra than methanol-free 
mixtures, and it takes into account the strong evidence of solid methanol in many protostellar environments \citep[e.g.][]{Pontopp_03_methdetect,GWB_04_ice}.
It is worth noting that, since good fits are obtained by both classes of models, neither method should be thought to provide a unique solution to the ice-composition 
analysis. 

\begin{figure}
 \begin{center}
 \includegraphics[width=0.8\textwidth]{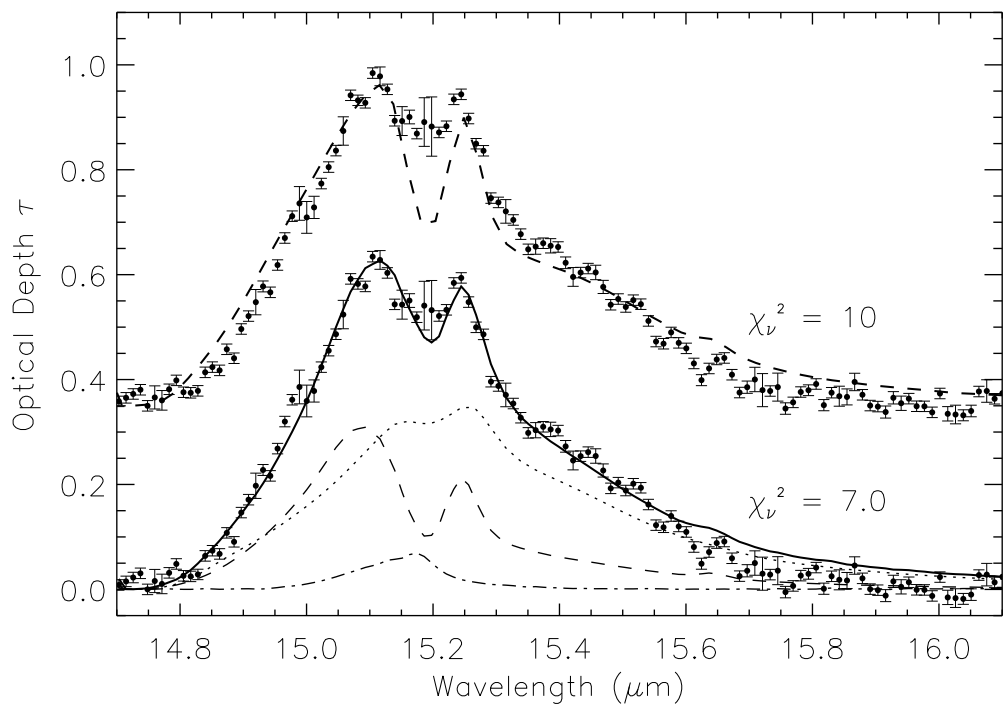}
 \end{center}
 \caption{15.2\mic feature of IRAS 04365+2535, shown repeated with a vertical offset.  The spectrum on the bottom is overlaid with a fit of 
H$_2$O:CH$_3$OH:CO$_2$ 0.9:0.3:1 at 98 K (dots), H$_2$O:CH$_3$OH:CO$_2$ 1:1:1 at 130 K (dashes), and CO:H$_2$O 10:1 at 30 K 
(dash-dot).  The solid line is the sum.  The heavy dashed fit on the upper spectrum is a fit using only H$_2$O:CH$_3$OH:CO$_2$ 1:1:1 at 118 K. 
The excess between 14.95 and 15\mic is due to gas-phase CO$_2$ ro-vibrational $\nu_2$ Q-branch lines, blended together at this resolution.}
 \label{fig:15umfit}
\end{figure}

Using a single component fit, we find that a nearly equal-parts mixture of H$_2$O:CH$_3$OH:CO$_2$ at temperatures of $\gtrsim$100 K provides the lowest-$\chi_\nu^2$ 
fit to the observed YSO 15.2\mic absorption bands, in accordance with previous results for more massive protostars \citep[e.g.][]{EKS_99_laboratory,GWE_99_molcloud}.  
Where present, the dual peak positions align, and the approximate strength of the 15.4\mic shoulder is replicated.  
The fit improves, however, by an average of $\sim$50\% of $\chi_\nu^2$, when we allow multiple components (Figure~\ref{fig:15umfit}).  The combination of these
additional components reproduces the relative peak strengths, the inter-peak dip depth, and the shape of the 15.4\mic shoulder without requiring a
specific single temperature.

Our multicomponent fits indicate
that for our YSO sample, a smaller amount of CH$_3$OH ($\sim$14\%) at lower temperatures is sufficient to account for the observed 15.4\mic shoulder, 
and equal parts H$_2$O, CO$_2$, and CH$_3$OH are reflected at high temperatures \citep[higher than those identified by e.g][]{EKS_99_laboratory}, 
when the CO$_2$ has at least partly migrated out of the matrix.
Laboratory mixtures at these higher temperatures that also contain the smaller 
fractions of CH$_3$OH have an asymmetrical dual peak that is not observed in the YSO spectra; 
mixtures with H$_2$O, CH$_3$OH, and CO$_2$ in a 1:1:1 ratio at lower temperatures show a stronger shoulder at 15.4\mic (relative to the main peak) as well as a 
nearly-single peak at 15.2 $\mu$m.  
While the cold-surface formation processes and expected abundances of CH$_3$OH are not tightly-constrained 
\citep[e.g.][and references therein]{Tielens_96_starformingices,KTB_01_5-8umregion}, the generally-higher
abundances observed towards high-mass protostars are evidence that warmer environments may be conducive to additional CH$_3$OH production.
This is consistent with the stronger 15.4\mic shoulders observed towards massive protostars 
\citep[compared with low-mass ones; e.g.][]{EDD_98_segregation,BPL_04_lowmass},
and the equal-parts H$_2$O:CH$_3$OH:CO$_2$ mixtures required to fit them.
Most of our YSO fits are further improved by including the very cold ($\sim$30 K) CO:H$_2$O layer, representing absorption 
along the line of sight due to cold material, affected by the region's radiation field but too distant or too shielded from the central source
to have undergone thermal heating.

\begin{figure}
 \begin{center}
 \includegraphics[width=0.8\textwidth]{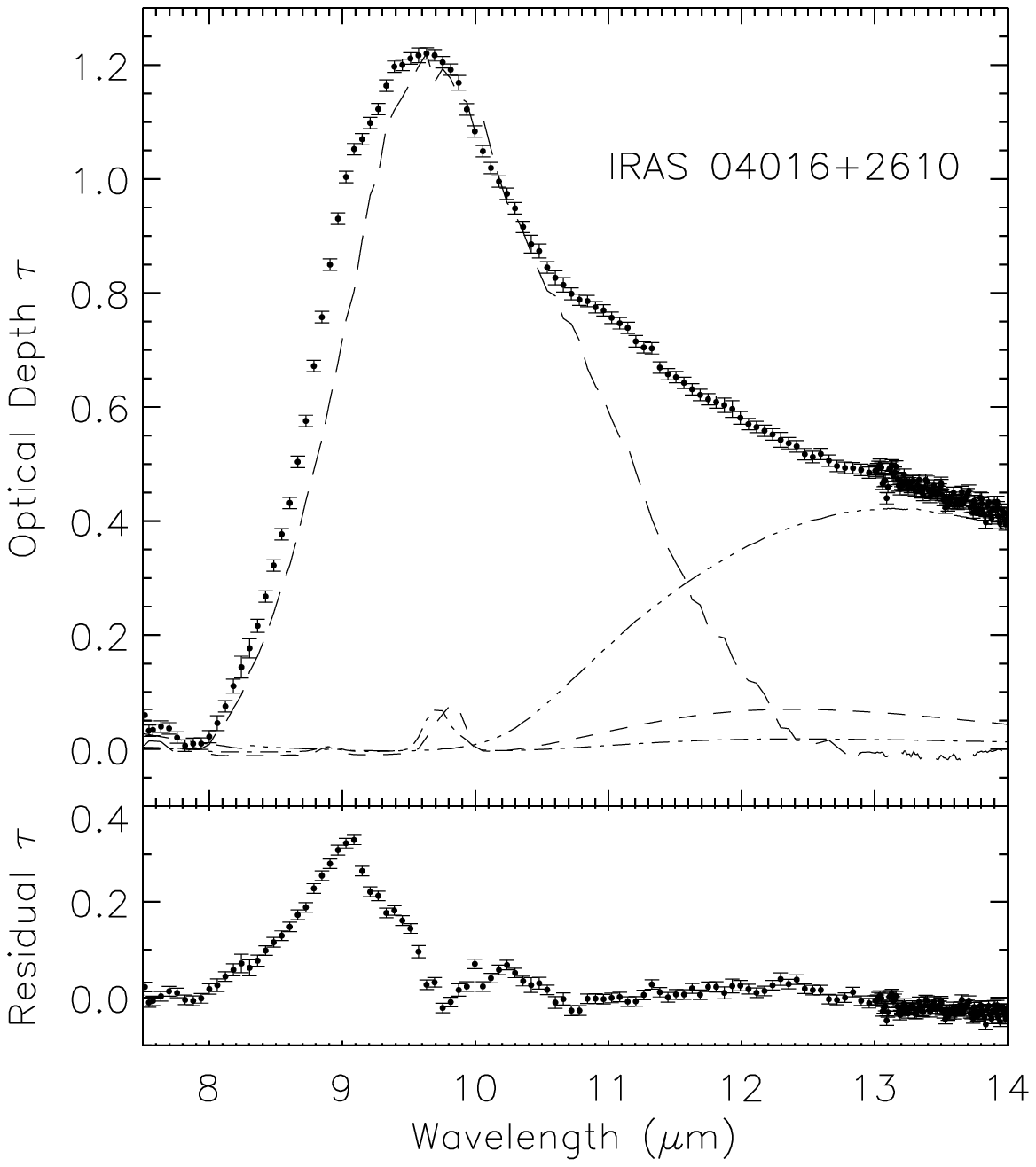}
 \end{center}
 \caption{Model of the absorption around 10\mic in IRAS 04016+2610.  Shown are the observed data (points), the GCS 3 silicate profile
\citep[long dashes;][]{Kemper_04_silicates}, pure water ice at 10 K (dash-triple dots), and the two methanol-bearing mixtures described in Section~\ref{sec:15um_feature}
and listed in Table~\ref{tab:15um_components} (short dashes and dot-dashes), scaled per this object's 15.2\mic fit.  The remnant peak at 9\mic is likely NH$_3$ ice.
The model and observation are consistent with one another in the relevant wavelength range.}
 \label{fig:silfit}
\end{figure}

Methanol-bearing ice mixtures have distinctive absorptions elsewhere in the mid-infrared spectrum, notably at 6.8 and 9.8 $\micron$. 
The former feature is discussed above (Section~\ref{sec:6.8um_feature}). 
The latter unfortunately lies in the very bottom of the silicate absorption feature. 
In the mixtures that best fit the 15.2 $\micron$ ice feature, the 
9.8 $\micron$ feature is either very weak or not particularly sharp. Thus it is difficult to separate possible 9.8 $\micron$ methanol ice absorption from small 
silicate-feature profile variation.
In general, a combination of absorption from the ice mixture that best fits the 15.2 $\micron$ feature, and from interstellar silicate grains 
\citep[toward GCS 3;][]{Kemper_04_silicates}, can be made to fit the spectrum, as illustrated in Figure~\ref{fig:silfit}. 
However, this result depends sensitively on the fit of 
the continuum and the silicate absorption profile, which as noted above is not very well-defined for these types of environments. We found no meaningful, 
quantitative constraints on a putative CH$_3$OH ice component from the 9.8 $\micron$ feature.

\begin{figure}
 \begin{center}
 \includegraphics[width=0.7\textwidth]{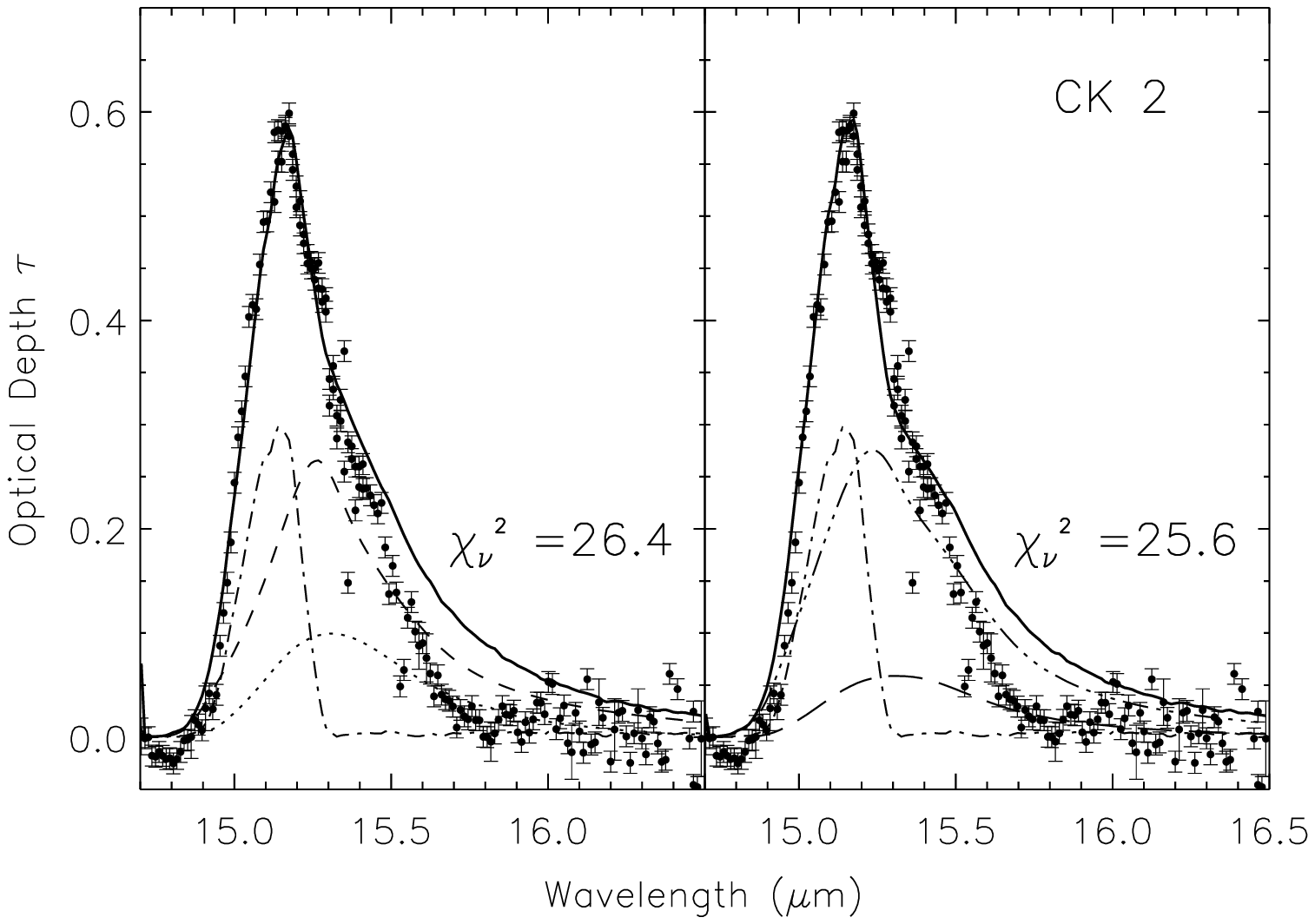}
 \includegraphics[width=0.7\textwidth]{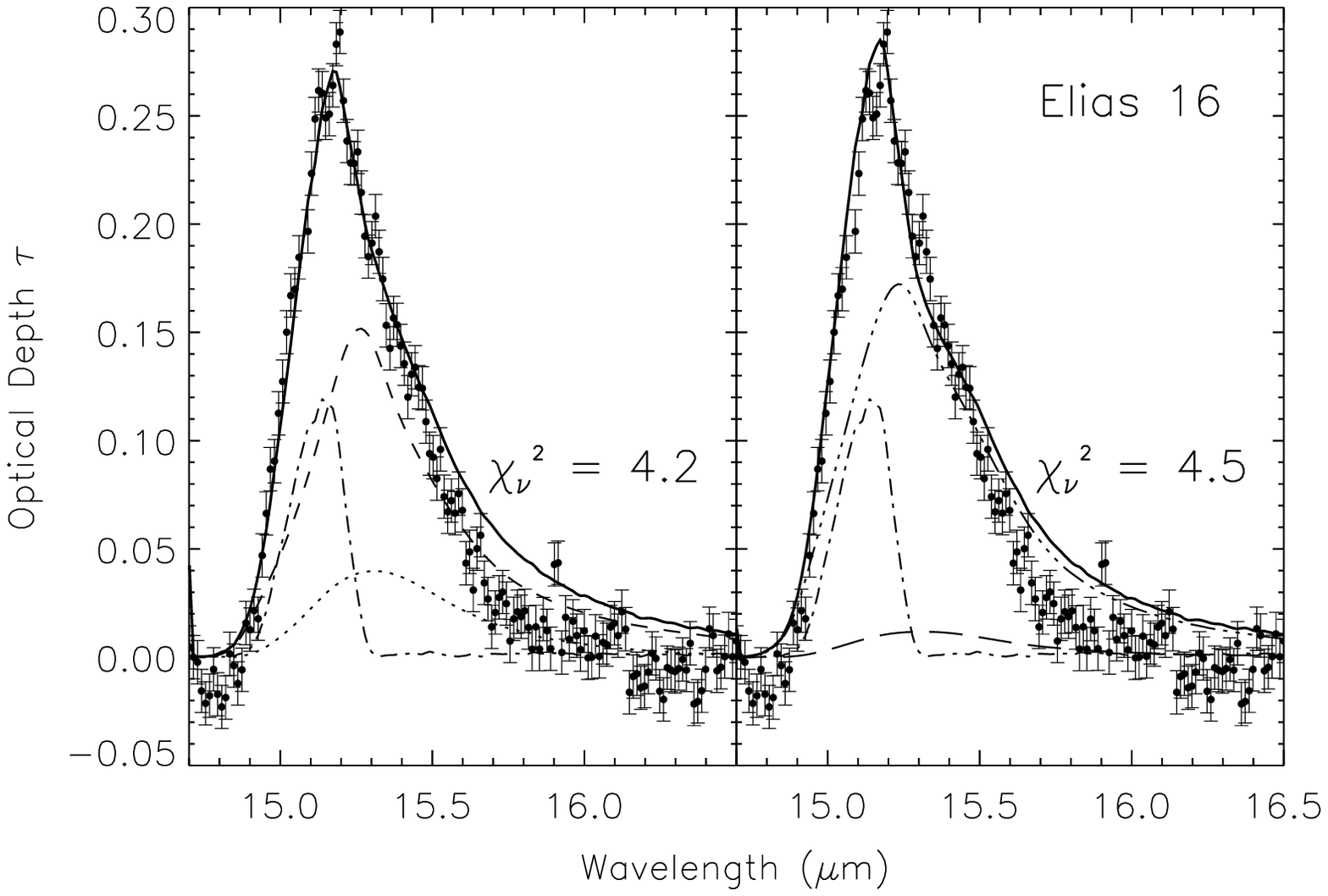}
 \end{center}
 \caption{Comparison of fits to the Serpens background star CK 2 (top) and Taurus background star Elias 16 (bottom).  
The left panels are repeats of the fits by \citet{KBP_05_backgroundstars}, where the
dotted line is a laboratory spectrum of H$_2$O:CO$_2$ 10:1 at 10 K, the dashed line is H$_2$O:CO$_2$ 1:1 at 10 K, and the dash-dot line is CO:N$_2$:CO$_2$ 10:5:2 at 30 K.
The
mismatch on the red wing is caused by the forced continuum at $\sim$16.5 $\mu$m, which ignores the extent of some of the ice feature wings as measured in the laboratory.
In the right panels, the dash-dot line is the same, but the H$_2$O:CO$_2$ 10:1 spectrum has been replaced with H$_2$O:CH$_3$OH:CO$_2$ 9:2:1 at 30 K (long dashes) and the
H$_2$O:CO$_2$ 1:1 with H$_2$O:CH$_3$OH:CO$_2$ 0.9:0.3:1 at 30 K (dash-triple dot).}
 \label{fig:bkgdfit}
\end{figure}

The 15.2\mic features in the molecular cloud sight-lines have previously been fitted with predominantly cold, polar H$_2$O:CO$_2$ mixtures
\citep{BMG_05_co2fieldstars,KBP_05_backgroundstars}.  We find that the $\chi^2_\nu$ of the fits can be matched or reduced
by including components containing methanol---specifically,
the cold counterparts of the H$_2$O:CH$_3$OH:CO$_2$ components used in our Class I/II fits described above (Figure~\ref{fig:bkgdfit}).  
The percentage of polar ice found ($\sim$75-80\%)
is comparable to that identified by \citet{BMG_05_co2fieldstars} and \citet{KBP_05_backgroundstars}, 
and the abundance ratios $N$(CO$_2$)/$N$(H$_2$O) from those studies are preserved.

We have calculated the total CO$_2$ column density (Table~\ref{tab:abundances}) observed along each embedded-object and molecular cloud
line of sight using Equation~\ref{eqn:column}. 
The exact band strengths of some of the laboratory mixtures used in the 15.2\mic fits
are undetermined.  The fitting results indicate that the vast majority of the CO$_2$ ice resides alongside H$_2$O ice,
so we use a band strength of 1.5$\times$10$^{-17}$cm molec$^{-1}$, 
which is appropriate for the 15.2\mic absorption from a H$_2$O:CO$_2$ 1.6:1 ice \citep{GSG_95_labsim}.  
The observed optical depth spectra were integrated between 14.7-16 $\mu$m.
We use the relative contributions of each fitting component to estimate the
column density of solid CO$_2$ in each stage of thermal processing (Table~\ref{tab:procseg}): radiation-only (10-30 K, ``cold''), some thermal heating (``warm''), 
and CO$_2$ segregation.  We have also estimated the column density of solid CH$_3$OH 
embedded in CO$_2$ using the the CH$_3$OH abundance and relative contribution of each laboratory fit component (Table~\ref{tab:abundances}).

\section{DISCUSSION} \label{sec:discussion}

\subsection{Ice Composition: Abundances With Respect to H$_2$O} \label{sec:chemdiscuss}

\subsubsection{CO$_2$ Column Density}

\begin{figure}
 \begin{center}
 \includegraphics[width=0.8\textwidth]{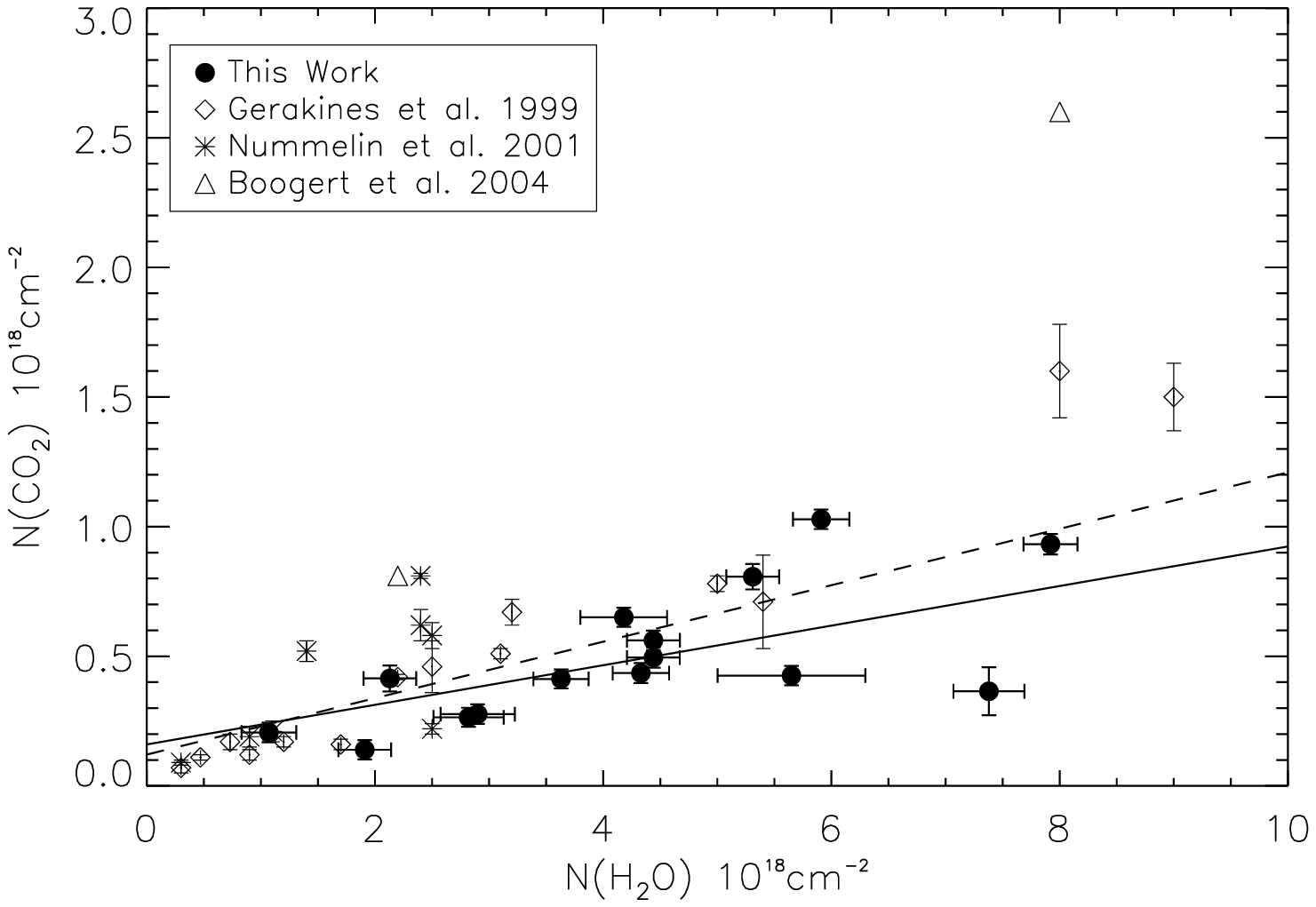}
 \end{center}
 \caption{Correlation between solid $N$(H$_2$O) and $N$(CO$_2$) for many different lines of sight.  
Our sample comprises low-mass protostars, \citet{GWE_99_molcloud} contains
mostly massive protostars and some galactic center sources, \citet{Nummelin_01_co2inlowmass} comprises low- and high-mass YSOs as well as some molecular cloud sight-lines,
and \citet{BPL_04_lowmass} contains two low-mass protostars.  The solid line is the ratio between column densities using only the current work's sample, 
and the dashed line is the ratio using all objects except the upper outlier from \citet{BPL_04_lowmass}.  Note that this work uses a CO$_2$ band
strength (at 15.2$\mu$m) $\sim$36\% higher than that of the previous studies, which may account for most of the difference in $N$(CO$_2$)/$N$(H$_2$O).}
 \label{fig:waterco2_Nratio}
\end{figure}

The column densities observed in our sample of Class I/II YSOs yield $N$(CO$_2$)/$N$(H$_2$O) $\sim$ 0.12($\pm$0.04) 
(Table~\ref{tab:abundances}, Figure~\ref{fig:waterco2_Nratio}), which is consistent with the values previously derived for many
different lines-of-sight, averaging $N$(CO$_2$)/$N$(H$_2$O) $\sim$ 0.17($\pm$0.03) 
\citep[][]{GWE_99_molcloud,Nummelin_01_co2inlowmass,GWB_04_ice,KBP_05_backgroundstars,ACA_03_isocam}, but significantly lower than the value of 
$N$(CO$_2$)/$N$(H$_2$O) $\sim$ 0.3, with large scatter, reported by \citet{BPL_04_lowmass} and 
\citet{Pontop_08_co2inlowmass}.  
Our study is unique in the sense that we use the band strength $A_{CO_2}$ of an H$_2$O:CO$_2$ mixture.  If we instead use the band strength for pure CO$_2$,
we find $N$(CO$_2$)/$N$(H$_2$O) $\sim$ 0.16($\pm$0.04), nearly equivalent to the results of the majority of previous studies, 
which include high- and low-mass YSOs, background field stars sampling molecular clouds, and galactic center sight-lines sampling 
diffuse interstellar material and molecular clouds.

The dashed line in Figure~\ref{fig:h2oco2_temp} identifies the locus of compositions with CO$_2$/H$_2$O = 0.12, which are clearly underrepresented in the heated samples
from the Leiden University database.  Good profile fits can only be achieved by combining ice mixtures, and this gap in composition-space may explain some of the
difficulty experienced in fitting 6\mic features.  Further laboratory exploration of the water-rich mixtures around the observed ratio with CO$_2$ may help
to improve constraints on the column density of H$_2$O ice based on the 6\mic feature.

\subsubsection{Evolutionary Trends in CO$_2$}

\begin{figure}
 \begin{center}
 \includegraphics[width=0.45\textwidth]{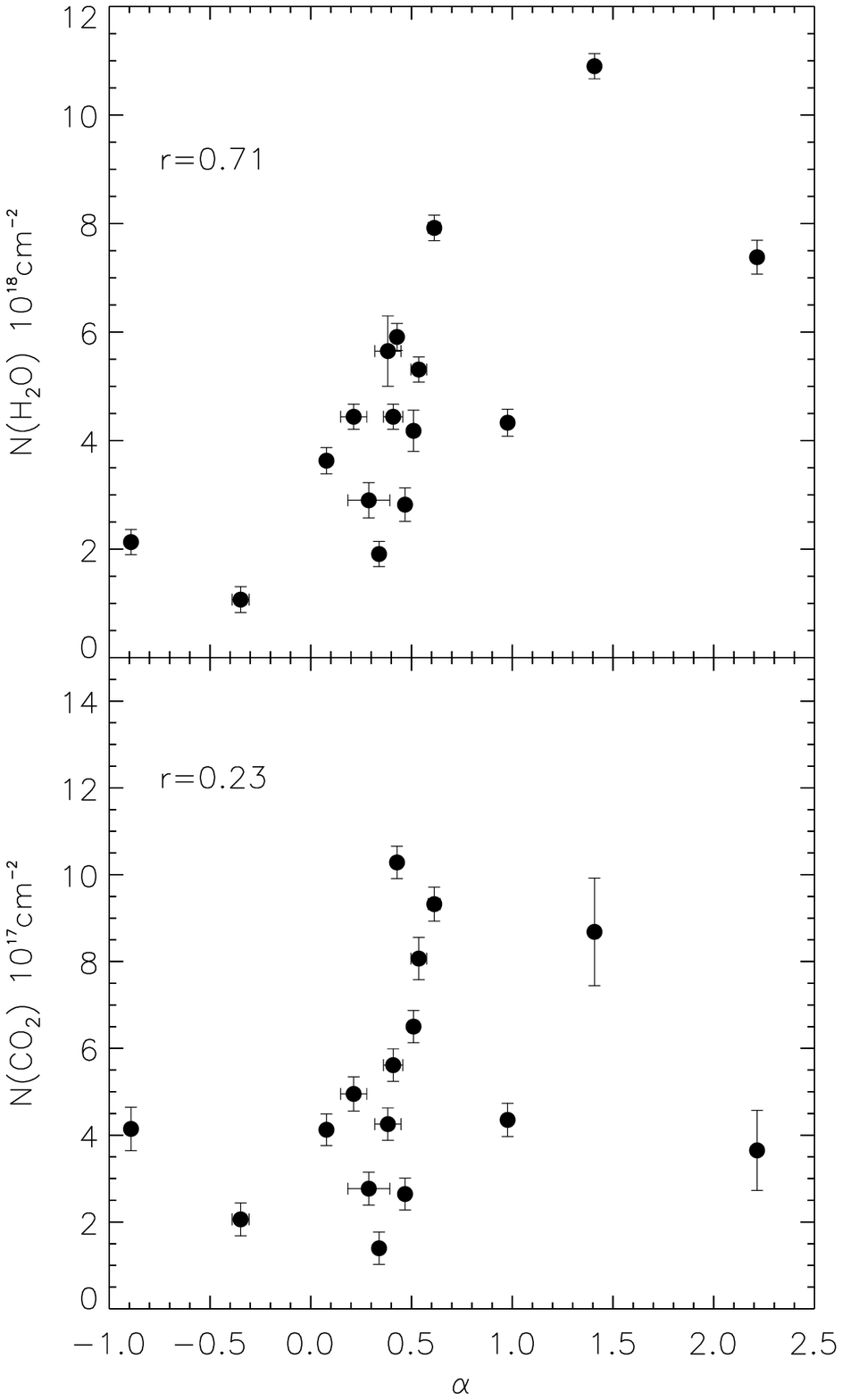}
 \includegraphics[width=0.45\textwidth]{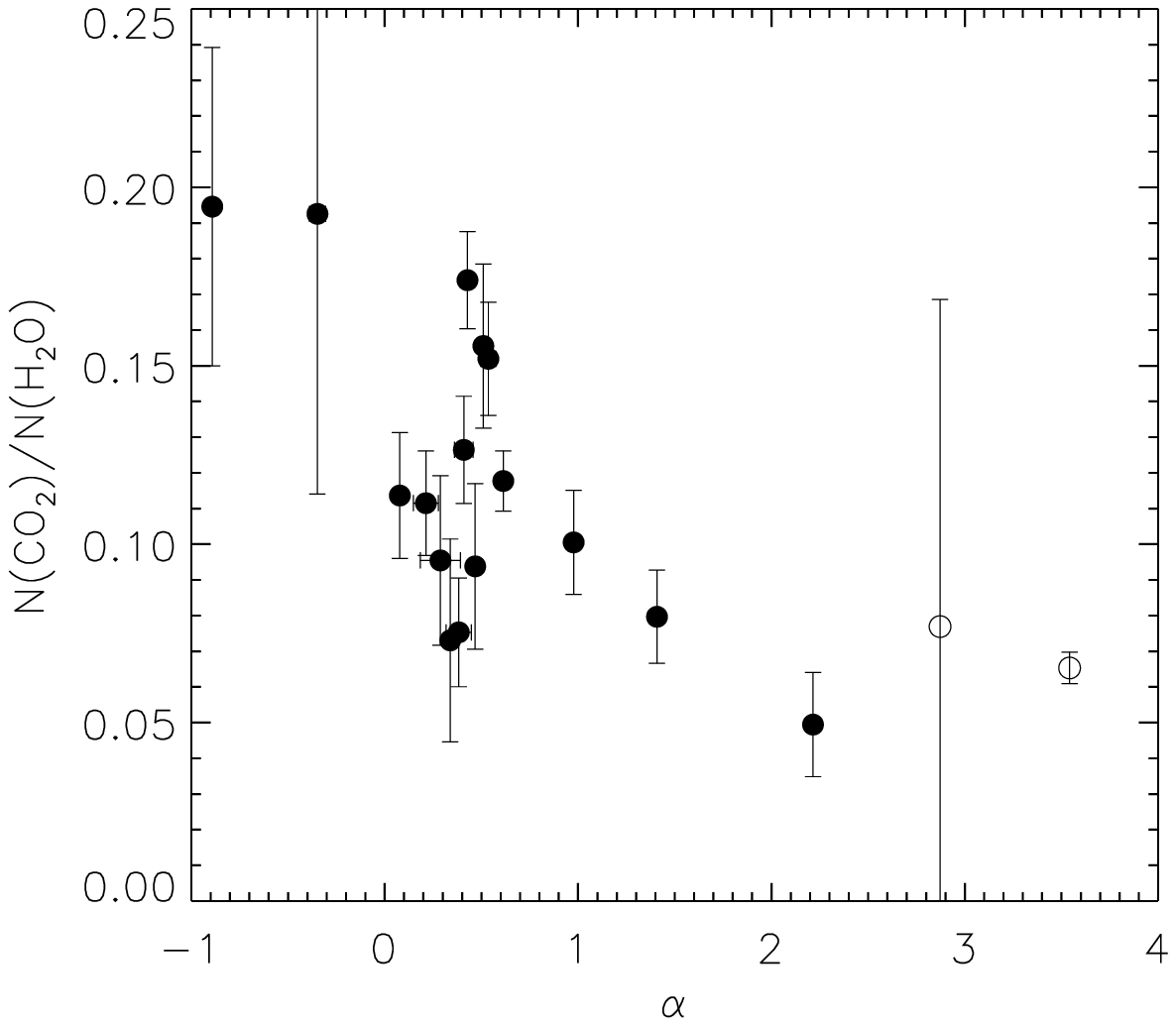}
 \end{center}
 \caption{Relationships between the SED spectral index ($\alpha$; 7-30 $\mu$m) and $N$(CO$_2$), $N$(H$_2$O), and $N$(CO$_2$)/$N$(H$_2$O).
The filled circles represent our Class I/II data, and the open circles represent the Class 0 objects.}
 \label{fig:alpha_abun}
\end{figure}

When considering the Class I/II sample in conjunction with the Class 0 objects, we find $N$(H$_2$O) to be well-correlated ($N$(CO$_2$) less so) with the
spectral index $\alpha$, where more embedded sources (higher $\alpha$) show higher column densities, as is expected (Figure~\ref{fig:alpha_abun}; top 2
panels).  The ratio $N$(CO$_2$)/$N$(H$_2$O) increases as the objects evolve towards lower $\alpha$-values (Figure~\ref{fig:alpha_abun}; bottom panel). 

The sublimation temperature of solid H$_2$O is $\sim$40 K higher than that of pure CO$_2$ ice \citep{GSG_95_labsim}.
Thus with increased exposure to thermal radiation, $N$(CO$_2$)/$N$(H$_2$O) is expected to go down, seemingly contradicting our observations. 
Although CO$_2$ molecules embedded in a polar matrix
may become ``trapped'' and able to survive to temperatures nearing the sublimation limit of H$_2$O, 
the thermal destruction of H$_2$O before CO$_2$ is unlikely.

An explanation of the high column density of CO$_2$ in more exposed environments may be found in the formation of additional CO$_2$ in
Class I/II objects.  
Photodissociation of H$_2$O leads to the production of radicals (e.g. OH) which can react to form CO$_2$ on the grain surface
\citep[e.g. CO+OH$\rightarrow$CO$_2$;][]{Alla_88_iceevolution}.
Some other reaction pathways to form CO$_2$ in molecular clouds and star-forming regions have activation barriers which become less obstructive in warmer,
more exposed environments \citep[e.g][]{Ruffle_01_co2formation,vanDishoeck_98_chemevolSFR}.  Indeed, in UV-irradiated H$_2$O:CH$_3$OH:CO$_2$ laboratory samples,
CO$_2$ eventually reaches a constant abundance level due to an equilibrium between destruction and formation \citep{EKS_99_laboratory},
while the abundance of H$_2$O further decreases due to photodissociation.

The values measured towards the background stars Elias 16, Elias 13, and CK~2 cannot be
extrapolated from this trend.  The $N$(CO$_2$)/$N$(H$_2$O) ratios for
these lines-of-sight (Table~\ref{tab:molcloud_abun}) are not lower than the ratios observed
towards the YSOs.  This discrepancy may be due to the fact that the
interstellar lines-of-sight probe a variety of environments and
compositional structure, while protostellar sight-lines are dominated
by the circumstellar material.  Possibly, the lower temperatures in
interstellar regions stifle CO$_2$ formation mechanisms with activation
barriers, while areas of increased density favor H$_2$O production.

\subsubsection{HCOOH}

\begin{figure}
 \begin{center}
 \includegraphics[width=0.8\textwidth]{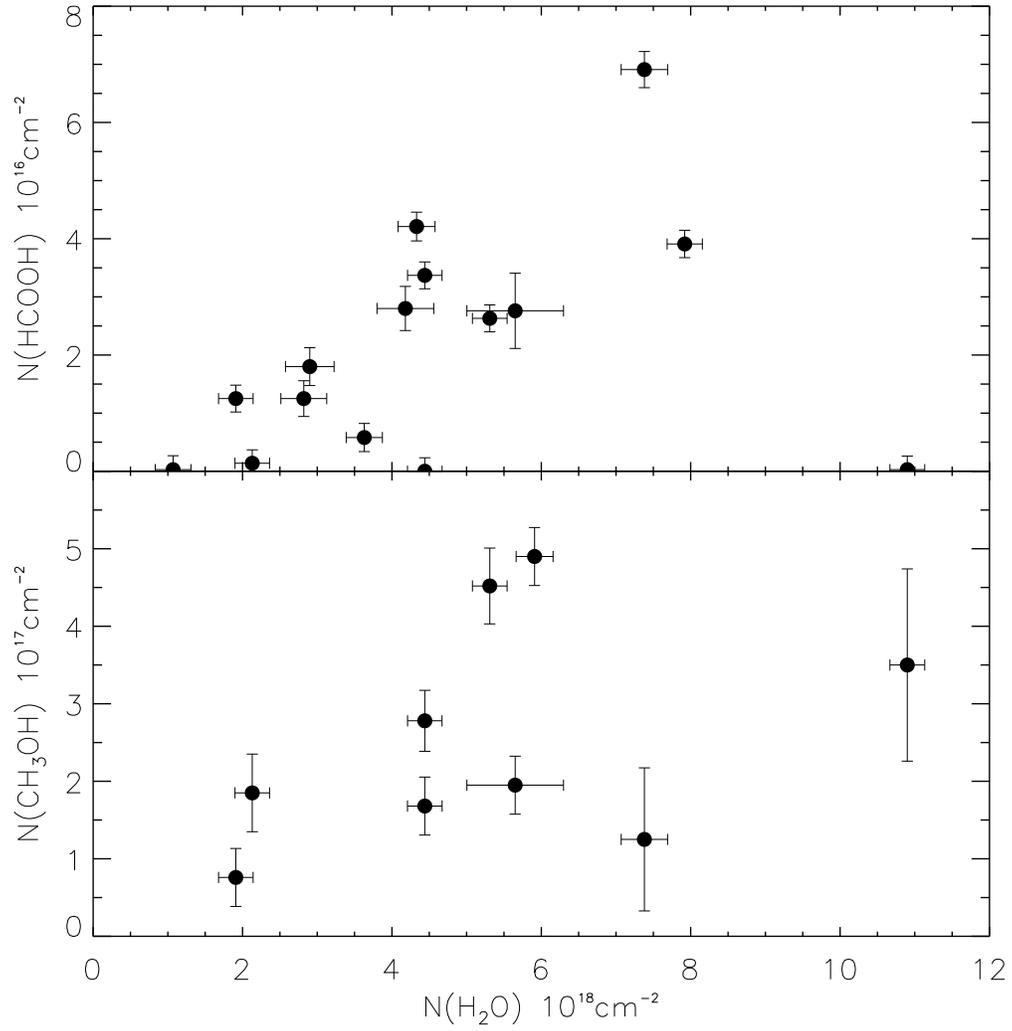}
 \end{center}
 \caption{Ratios between $N$(H$_2$O), and $N$(HCOOH) and $N$(CH$_3$OH), for the Class I/II sample (Section~\ref{sec:chemdiscuss}).  The bottom plot refers to 
$N$(CH$_3$OH) embedded in a H$_2$O:CO$_2$ matrix.}
 \label{fig:waterhcoohmeth_Nratio}
\end{figure}

Apart from one outlier at $N$(H$_2$O) $\sim$ 1.1$\times$10$^{19}$cm$^{-2}$ (L1551 IRS 5), the column densities of HCOOH and H$_2$O are strongly related ($r=0.61$,
Figure~\ref{fig:waterhcoohmeth_Nratio}).  This supports laboratory experiments demonstrating that the most likely formation environments
of solid HCOOH are those that also produce H$_2$O ice, through either radiative processing or cold surface reactions \citep{Bisschop_07_hcooh}.
We find $N$(HCOOH)/$N$(H$_2$O) $\sim$ 0-2\%, with an average of 0.6\%, 
well within the $\lesssim$5\% abundance ratio observed in lines-of-sight towards massive and low-mass protostars \citep{Bisschop_07_hcooh,Boogert_08_lowmassices}.

\subsubsection{CH$_3$OH}
The observed $N$(CH$_3$OH)/$N$(H$_2$O) ratios in our sample do not show a clear trend (Figure~\ref{fig:waterhcoohmeth_Nratio}).  
It is known that the $N$(CH$_3$OH)/$N$(H$_2$O) abundance varies greatly from source to source \citep[e.g][]{BE_04_ISI}, and it may be better
to analyze the CH$_3$OH abundance with respect to CO$_2$.

\begin{figure}
 \begin{center}
 \includegraphics[]{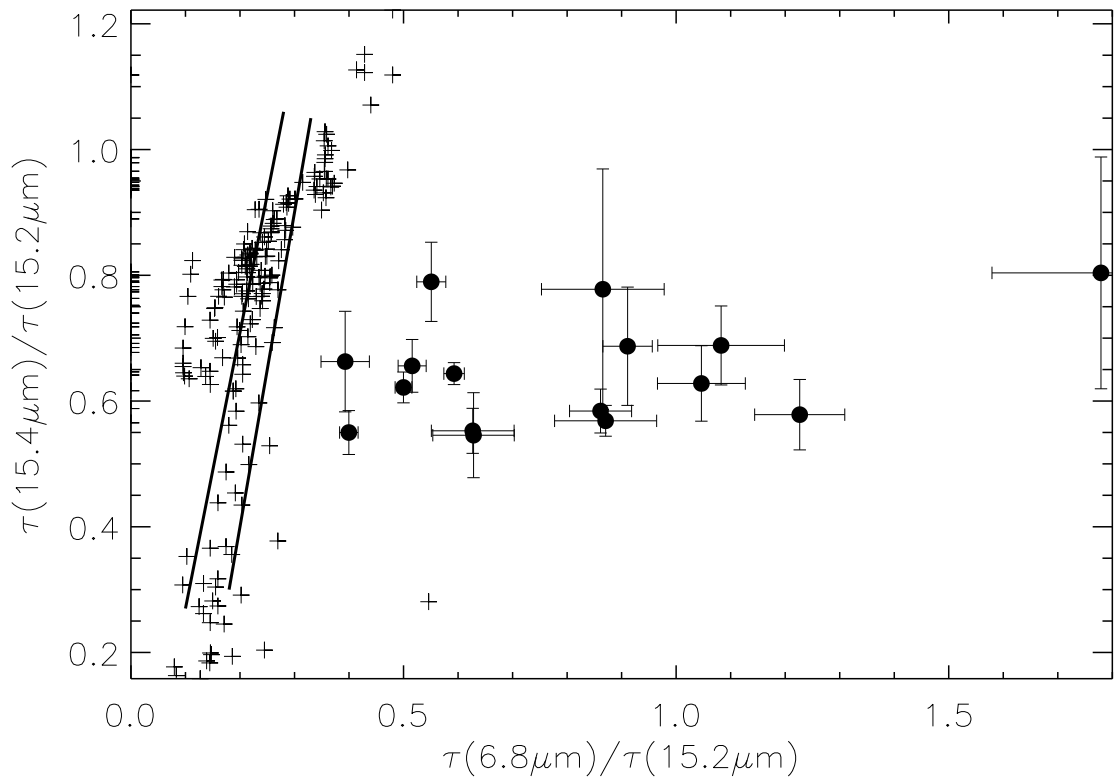}
 \end{center}
 \caption{The correlation between 6.8\mic and 15.4\mic optical depths in the laboratory ice spectra (plus signs) and the observed YSO spectra (solid circles).
Both have been scaled by peak 15.2 $\mu$m optical depth.  The laboratory data cover temperatures from 10-185K and CH$_3$OH/CO$_2$ abundance ratios of 0-1.4, which are 
roughly indicated by the solid lines: [$\lesssim$0.5] {\bf /} [0.5-1.4] {\bf /} [$\gtrsim$1.4].  The vertical spread depends largely on temperature, with warmer mixtures
of a constant composition showing weaker 15.4\mic shoulders.}
 \label{fig:6.8um15.2um_vs_15.4um15.2um}
\end{figure}

CH$_3$OH can be detected in the spectrum in three places. When it is intermixed with CO$_2$, a shoulder at 15.4 $\mu$m due to CH$_3$OH can
be observed in the 15.2 $\mu$m feature; and CH$_3$OH also shows 6.8 and 9.8 $\mu$m bands (the latter is blended in the 9.7\mic silicate complex).
In Figure~\ref{fig:6.8um15.2um_vs_15.4um15.2um} we show the ratio of the optical depth at
15.4 $\mu$m (vertical axis) and 6.8 $\mu$m (horizontal axis), both with respect to the optical depth at 15.2 $\mu$m.  Filled circles
represent the astronomical measurements, while the plus signs indicate the ratios measured in the laboratory spectra.  The laboratory spectra
show a correlation between the two features, whose exact position depends on the temperature.  The 15.4/15.2 $\mu$m $\tau$ ratios
observed in the astronomical data are within the range measured in the laboratory, but the corresponding 6.8 $\mu$m optical depths are
significantly greater than the laboratory values.  This could be understood if only a small fraction of the CH$_3$OH along the line of
sight is embedded in a CO$_2$ matrix or if there are significant contributions to the 6.8 $\mu$m feature from other components.

Using the laboratory fits to the 15.2\mic complex in the high spectral resolution data, we estimate the column density of solid CH$_3$OH intermixed with CO$_2$
as 30-56\% with respect to CO$_2$, corresponding to $N$(CH$_3$OH)$_\mathrm{{CO_2}}$/$N$(H$_2$O) $\sim$ 2-9\% (Table~\ref{tab:abundances}).
Assuming that $\lesssim$50\% of the 6.8\mic feature is due to to CH$_3$OH (Section~\ref{sec:6.8micfits}), we estimate $N$(CH$_3$OH)$_\mathrm{{total}}$/$N$(H$_2$O) 
$\lesssim$ 5-17\%.  There are two outlying values, for IRAS 04108+2803A and IRAS 04181+2654B, at 32 and 28\%.  
Although we do not place high confidence in these as absolute values because of the 6.8\mic uncertainties, they are
consistent with the detection of solid CH$_3$OH with abundances (relative to H$_2$O) in this range up to $\sim$25\% towards low-mass YSOs \citep[as well as the overall
scatter in abundance;][]{Pontopp_03_methdetect,Boogert_08_lowmassices}.  Thus our data support the presence of a significant CO$_2$-poor H$_2$O:CH$_3$OH ice component.

\subsubsection{CH$_4$}

\begin{figure}
 \begin{center}
 \includegraphics[width=0.8\textwidth]{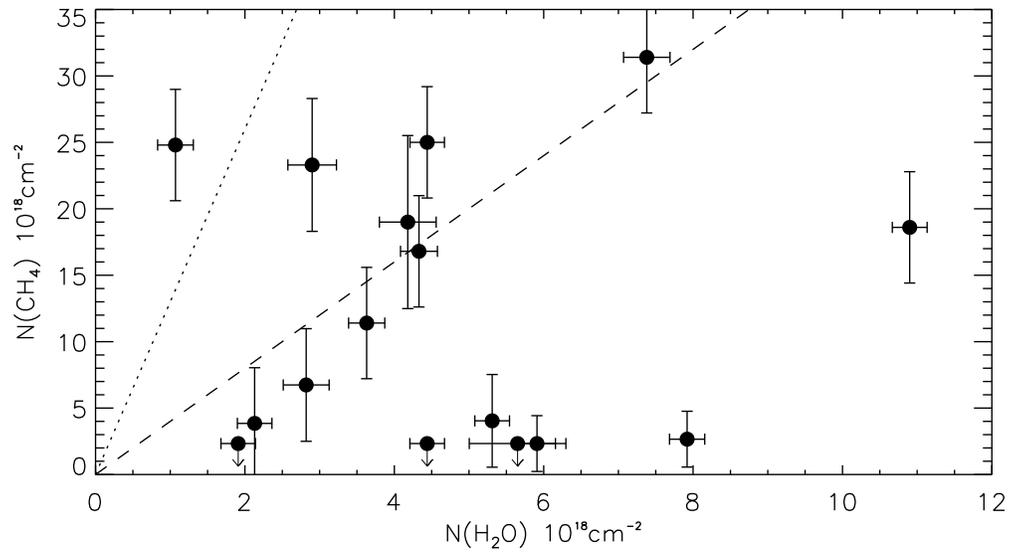}
 \end{center}
 \caption{Comparison of solid $N$(H$_2$O) and $N$(CH$_4$) for the YSO sample.  
The dashed line indicates the maximum of $\sim$4\% (CH$_4$ relative to H$_2$O) observed towards massive protostars \citep[e.g.][]{GWB_04_ice}, and the 
dotted line is the newer 13\% maximum reported by \citet{Oberg_08_ch4} towards low-mass protostars.}
 \label{fig:watermethane_Nratio}
\end{figure}

We observe a range of $N$(CH$_4$)/$N$(H$_2$O) ratios from $\sim$0.3 to 8\% in our sample (Figure~\ref{fig:watermethane_Nratio}), 
with an average of 2.6\% and a single outlier (IRAS 04154+2823) at 23\% .  
Our measurements largely overlap with the ratios of $\sim$0.5-4\% reported in the literature \citep[e.g.][]{Boogert_96_ch4inprotostars,GWB_04_ice},
though ratios up to 13\% are reported by \citet{Oberg_08_ch4}.
The higher values in our sample ($>$4\%, dashed line in Figure~\ref{fig:watermethane_Nratio}) may be 
partially ascribed to additional absorptions unresolvable at the SL-module resolution,
such as gas-phase CH$_4$ lines \citep{Boogert_97_labch4}, or to matrix-dependent variations in band strength.  
\citet{Hudgins_93_ch4bandstrength} found that the band strength for the 7.7\mic
CH$_4$ feature may vary by as much as a factor of 2.8 between H$_2$O-rich and CO-rich matrices.  
We use the value for pure CH$_4$, which lies approximately midway between the two extremes.

Laboratory studies of solid CH$_4$ indicate that it can be formed by a variety of processes---including UV photolysis of CH$_3$OH ices \citep{Alla_88_iceevolution}
and grain-surface hydrogenation of carbon \citep{Tielens_82_compositionmodels}---and that the resulting absorption profile
has characteristics unique to its formation pathway and embedding matrix \citep{Boogert_97_labch4}.  
Unfortunately, the low signal-to-noise of the CH$_4$ detections, and the low spectral resolution, prevent a detailed analysis of the spectral profiles;
however, the broad range of peak positions and widths implies that the CH$_4$ embedding matrices vary across the sample.

\subsection{Segregation of CO$_2$ Ice} \label{sec:co2seg}

\begin{figure}
 \begin{center}
 \includegraphics[width=0.4\textwidth,angle=90]{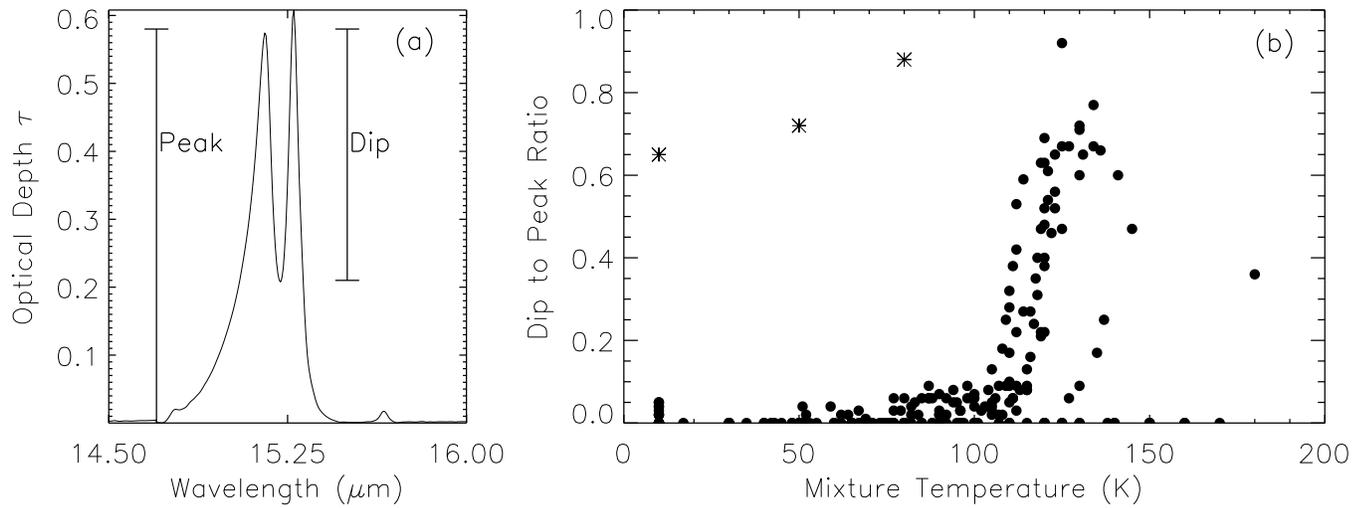}
 \end{center}
 \caption{The ``dip-to-peak'' ratio as a measure of the amount of CO$_2$ segregation.  {\it (a)}---Demonstrates how the values were measured for each laboratory ice
spectrum, using pure CO$_2$ as an example.  
{\it (b)}---The relative dip strength as a function of laboratory ice temperature, showing the rapid onset of the process near $\sim$105 K.
Much of the scatter in (b) is due to the dependence of CO$_2$ segregation temperature on the exact composition of the matrix in which it is embedded.  
The large asterisks in the upper left represent pure CO$_2$ ice.}
 \label{fig:d2p_vs_temp}
\end{figure}

In the colder parts ($\lesssim$20 K) of a star-forming region, well-removed from the hot cores and embedded protostars or buried in dense clumps, 
apolar CO$_2$ ice is formed embedded in a polar H$_2$O matrix \citep[e.g.][]{BE_04_ISI}.
The 15.2\mic CO$_2$ resonance will appear to be broad and single-peaked.  As these ices are heated,
as for instance happens in the circumstellar environments of YSOs,
the CO$_2$ can migrate out of the polar matrix and aggregate, 
either as a surface layer or in bulk apolar clumps within the matrix \citep[e.g.][]{SA_90_co2inanalogs,EKS_99_laboratory}.
The spectral appearance of the CO$_2$ resonance changes, and a double-peaked structure, with peaks at 15.1 and 15.25 $\mu$m, becomes apparent rather
rapidly (Figure \ref{fig:d2p_vs_temp}).
The migration process also destabilizes the CO$_2$:CH$_3$OH interactions
and reduces the 15.4\mic shoulder.  Due to the apparent spectral difference between segregated CO$_2$ ice
and CO$_2$ ice embedded in an H$_2$O:CH$_3$OH matrix, it is possible to determine the relative mass contained in both of these phases.

\begin{figure}
 \begin{center}
 \includegraphics[width=0.7\textwidth]{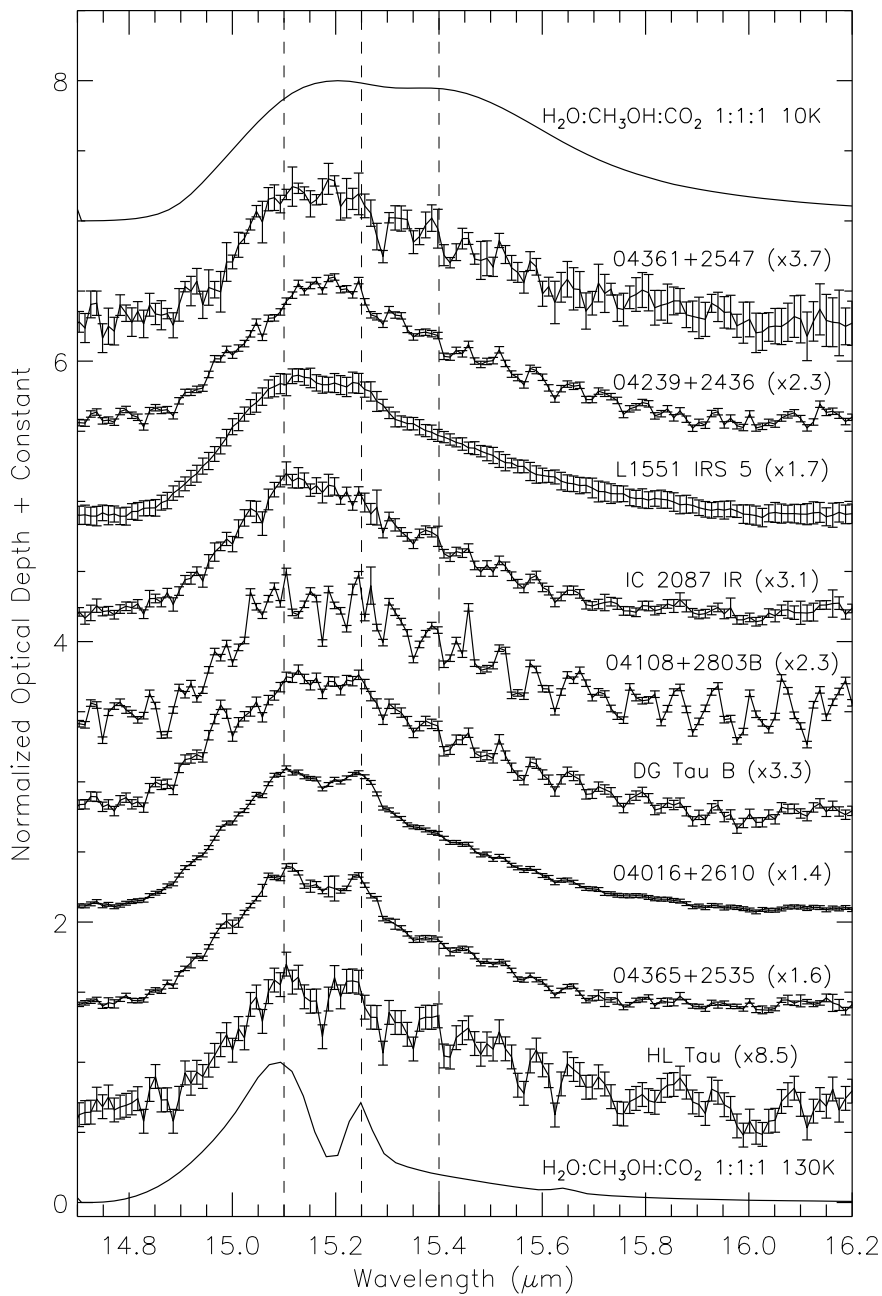}
 \end{center}
 \caption{The high-resolution 15\mic Class I/II profiles visually ordered by ``peakiness,'' normalized to 1, and vertically offset for clarity. 
The dotted lines at 15.1 and 15.25\mic mark the peak positions of pure CO$_2$, and the one
at 15.4\mic indicates the CH$_3$OH:CO$_2$-complex shoulder.  The top and bottom spectra are laboratory mixtures as indicated.}
 \label{fig:15um_hires_spectra}
\end{figure}

In Figure~\ref{fig:15um_hires_spectra}, we show the variation in the CO$_2$ 15.2 $\mu$m absorption feature in our sample, along with laboratory spectra of
CO$_2$ fully embedded in the matrix (top; 10 K) and fully segregated CO$_2$ (bottom; 130 K). The dashed lines indicate the positions of the
double-peaked segregated CO$_2$ and the shoulder at 15.4 $\mu$m due to embedded CO$_2$ interacting with CH$_3$OH. We show
only the Class I/II objects for which high resolution spectra are available, as the double-peaked structure is not resolved
in the low resolution spectra. 
Several of the spectra clearly show the double-peaked structure, while all deviate from the smooth, unprocesed Taurus cloud profiles
as published by \citet{WSB_07_quiescentICM}, \citet{BMG_05_co2fieldstars}, and \citet{KBP_05_backgroundstars}.
Indeed, from the multi-component fits described in Section~\ref{sec:15um_feature},
we find evidence in all of our low-mass YSO spectra for moderate-temperature thermal processing 
($\lesssim$100 K; CO$_2$ warmed but not segregated),
in agreement with \citet{BPL_04_lowmass} but in contrast with \citet{Nummelin_01_co2inlowmass}.
However, unlike \citet{BPL_04_lowmass}, we also find
evidence for high-temperature processing---5 of the 9 objects (IRAS 04016+2610, IRAS 04239+2436, IRAS 04365+2535, DG Tau B, and IC 2087 IR)
have 15.2\mic profiles entirely dominated by warmed ices in which a significant fraction ($\sim$20-40\%)
of the CO$_2$ has segregated from the polar formation matrix.  

From the multi-component fits, we can derive the mass distribution of CO$_2$ ice over the {\it phases} (cold, warm, segregated).
The phase transitions correspond to laboratory temperatures, which in turn translate to (lower) astrophysical temperatures.
The onset of CO$_2$ segregation occurs at a temperature near 105 K in the laboratory (Figure \ref{fig:d2p_vs_temp}), which corresponds
to an astrophysical temperature of $\sim$65 K \citep[e.g.][]{BEG_00_13co2}.
The CO$_2$ evaporation temperature depends on the relative abundance of other 
species, particularly H$_2$O, but the mixtures with the most physically likely abundances 
(i.e. CO$_2$/H$_2$O $<$ 1) show an average sublimation temperature of $\sim$150 K, or 80 K in astrophysical environments.
Thus, the {\it segregated} column densities of Table~\ref{tab:procseg} represent CO$_2$ ice at circumstellar radii where the
ice temperature is between 65 and 80 K, and the {\it cold}+{\it warm} column densities indicate ice farther from the central source (i.e. where T $<$ 65 K).
The inner boundary of the segregated CO$_2$ ice zone is defined by the evaporation temperature at $\sim$80 K.
The low-mass protostellar envelope models produced by \citet{Furlan_08_classImodel} for nearly all the stars in our sample 
indicate these approximate temperature ranges in the densest inner regions
($R$ $\lesssim$ 100 AU); however, detailed physical structure models that also include the appropriate ice opacities are needed to further constrain the location and
processing levels of the ices.

\begin{figure}
 \begin{center}
  \includegraphics[width=0.6\textwidth]{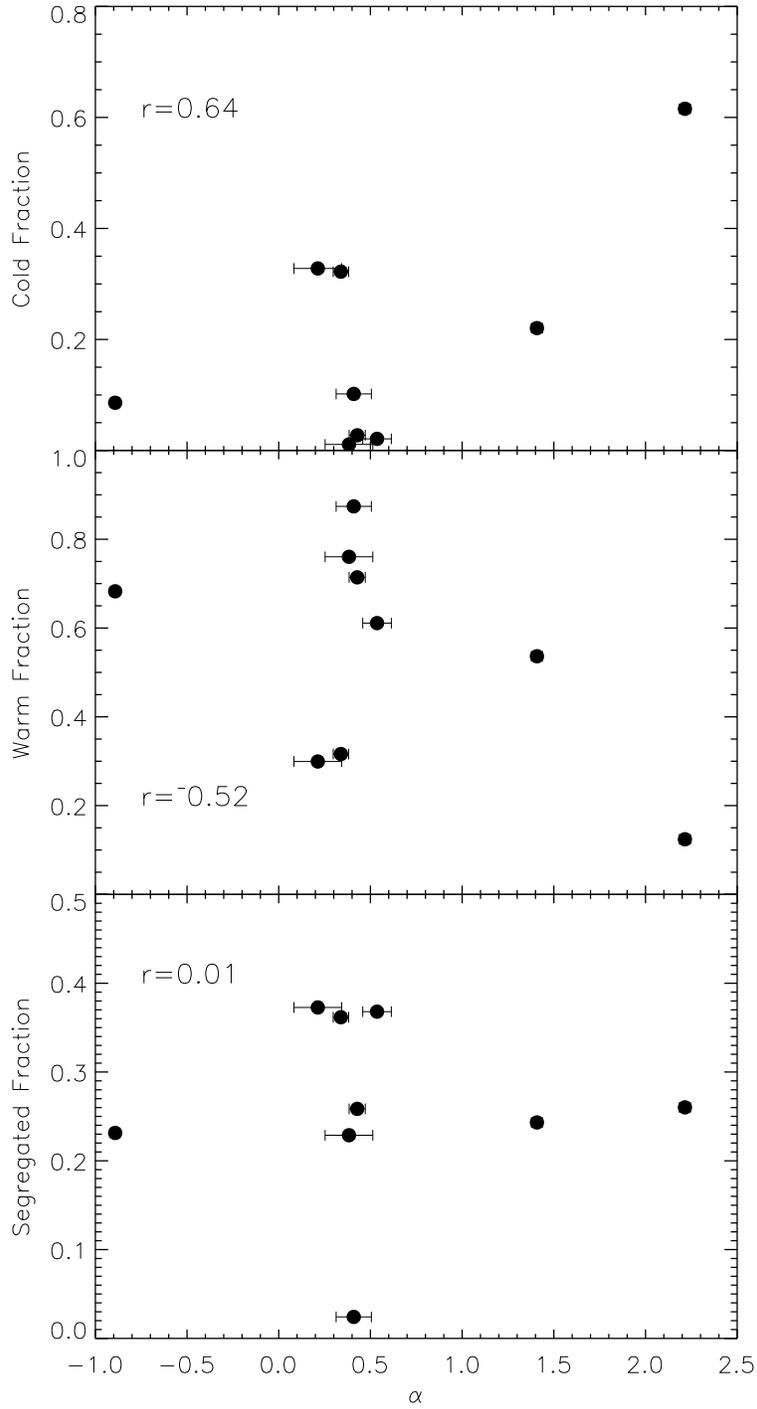}
 \end{center}
 \caption{Relationship between the SED spectral index ($\alpha$; 7-30 $\mu$m) and the CO$_2$ processing fractions in our Class I/II sample.}
 \label{fig:alpha_seg}
\end{figure}

In Figure~\ref{fig:alpha_seg} we show the correlation of thermal processing levels with the SED spectral index (indicative of the line-of-sight dust column).  
While there is a large amount of scatter in these plots, the overall fraction of cold CO$_2$ ice increases 
(and the warm fraction decreases) for more deeply embedded objects, while the segregated fraction does not show a clear trend with total dust opacity.

\subsubsection{Segregation of CO$_2$ Ice in Molecular Clouds and W3(OH)}
All three interstellar cloud sight-lines show
single-peaked 15.2\mic profiles with very little substructure; CK 2 and Elias 16 are remarkably similar in shape (Figure~\ref{fig:bkgdfit}), 
despite indications that ice mantles in
Serpens may be fundamentally different from those in Taurus \citep[e.g][]{Eiroa_89_serpensice,WSB_07_quiescentICM}.  
These results give support for the cold grain origin of at
least some of the CH$_3$OH observed towards protostars; the presence of the unidentified 6.8\mic band in background star sight-lines
\citep[in particular CK 2 and Elias 16;][]{KBP_05_backgroundstars} is consistent with the assignment of CH$_3$OH as a carrier.

\begin{figure}
 \begin{center}
 \includegraphics[width=0.7\textwidth]{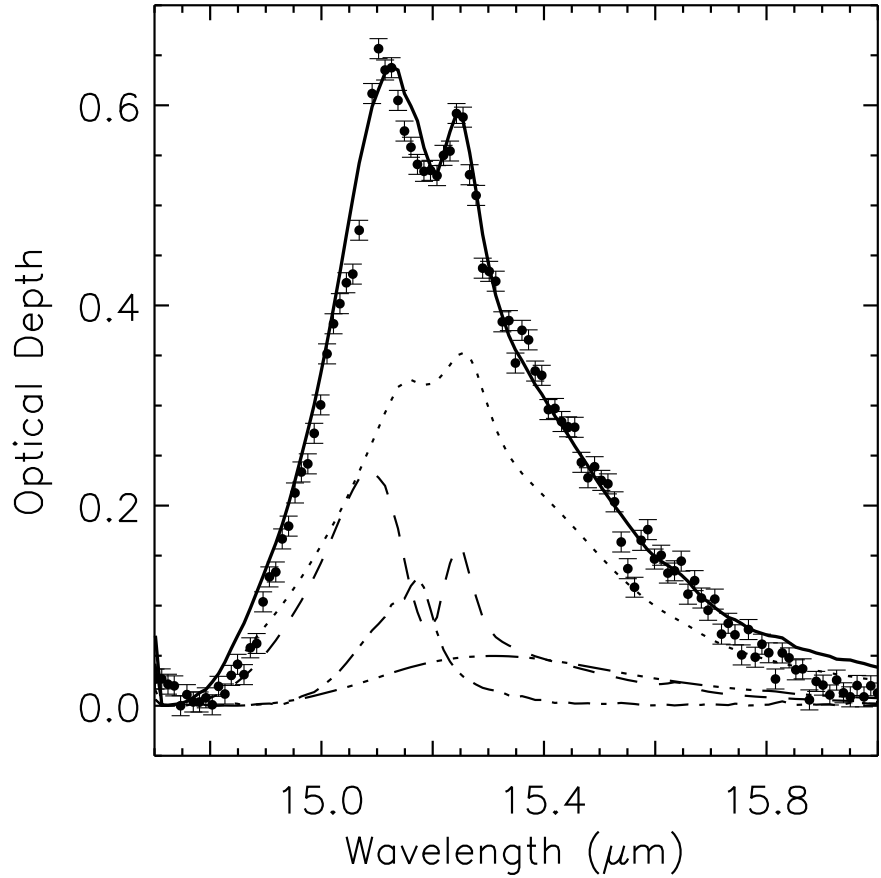}
 \end{center}
 \caption{15.2\mic CO$_2$ feature probing W3(OH) and the foreground W3 cloud.  
It has been fitted with a combination of H$_2$O:CH$_3$OH:CO$_2$ 0.9:0.3:0.1 at 98 K (dotted line),
H$_2$O:CH$_3$OH:CO$_2$ 1:1:1 at 130 K (dashed line), CO:H$_2$O at 10:1 at 30 K (dot-dashed line), and H$_2$O:CO$_2$ 10:1 at 10 K (dash-triple dot).  
The solid line is the sum.}
 \label{fig:w3ohfit}
\end{figure}

Interestingly, the spectrum of W3(OH), which has the reddest SED of our entire sample, shows a clearly double-peaked CO$_2$ feature, 
indicating pure CO$_2$ along the line of sight, which could have been formed by thermal segregation (Figure~\ref{fig:w3ohfit}).  
The segregation fraction is $\sim$26\%, comparable to the Class I/II average of $\sim$30\%.  Given the difference in segregation with 
the pure molecular cloud lines of sight, the most likely
explanation is that this line of sight samples not only the cold foreground W3 cloud but also the warmer, 
thermally-processed CO$_2$ ices around the massive protostars in
the W3(OH) region.  The strong similarity of the 15.2\mic absorption profile towards W3(OH) to those of the low-mass Taurus
objects is curious, as the line of sight toward W3(OH) is highly diverse, containing an ultra-compact H{\sc ii} region,
strong outflows, and H$_2$O/OH maser emissions \citep{Turner_84_w3ohtw,Alcolea_93_whohtwmaser,Argon_03_w3ohtwmaser}.  
Future spectroscopic observations with higher angular resolution will resolve the spatial structure and
provide useful information on the complexity of the early stages of CO$_2$ heating in embedded objects.

\section{CONCLUSIONS} \label{sec:conclusions}

We have analyzed the mid-infrared molecular ice features in 16 Class I/II low-mass YSOs using Spitzer IRS observations.  
Using detailed comparisons to laboratory ice analog spectra, we find the ice mantle compositions to be dominated by H$_2$O and CO$_2$ ice but with significant
abundances of additional species such as CH$_3$OH, H$_2$CO, HCOOH, NH$_3$, CH$_4$, and SO$_2$ ($\lesssim$10\% each).  

Detailed spectral fits to the 6\mic feature reveal that $\gtrsim$80\% of the band can be attributed to H$_2$O, with additional absorptions near 5.85\mic identified as
HCOOH and H$_2$CO.  Some of the observed spectra have a broader 6\mic peak that can be well-fit by H$_2$O:NH$_3$ ices, 
providing evidence for the presence of solid NH$_3$.  
On the long-wavelength wing of the 6\mic feature, small absorption remnants are observed, varying
in strength, shape, and peak position among the sources.  
We show that these are highly unlikely to be the result of PAH absorption, 
due to the absence of other expected PAH features at 7.5-9 and 11-14 $\mu$m.

We have examined the unidentified 6.8\mic feature and discussed its assignment to solid CH$_3$OH and NH$_4^+$, the two most common candidates.  While no high-quality
spectral fits could be achieved using these two components, we find that on average the observed YSO features could be approximately fit by equal contributions
from NH$_4^+$ and from a CH$_3$OH/H$_2$O$\sim$0.1 mixture, with varying amounts of nearly-pure CH$_3$OH intermixed.  However, we find a wide variety
of feature shapes in our sample, indicating that the interstellar component(s) producing this band vary greatly from object to object, and a single simple
carrier or set of carriers may be inadequate to explain the feature.

In the 7-8\mic range, we are able to isolate and identify features attributed to CH$_4$, SO$_2$, and
(tentatively) HCOOH ices.  The high CH$_4$ ice abundances derived from these, along with the variety of feature shapes and peak positions within the sample, 
suggest variations in line-of-sight molecular environment or possible absorption contamination by unresolvable gas-phase CH$_4$ lines.  
The solid SO$_2$ and CH$_4$ band profiles are
very sensitive to molecular environment and temperature, and additional high-resolution studies of this wavelength range will yield further information
on molecular ice interactions and abundance structures along the line of sight.

The 15.2\mic solid CO$_2$ bending mode feature is well-reproduced by multi-component laboratory fits comprising ice analogs (H$_2$O, CH$_3$OH, CO, CO$_2$)
in various stages of radiative and 
thermal processing.  Using the relative contribution of the components to each high-resolution feature profile, we find that all of the YSO ice environments
have undergone at least some thermal heating, and 5 of the 9 have profiles entirely dominated by processed ices in which a significant fraction ($\sim$20-40\%)
of the CO$_2$ has segregated from the polar formation matrix.  

The molecular ice abundance ratios, particularly $N$(CO$_2$)/$N$(H$_2$O), are consistent with the values observed in a wide variety of sight-lines, but within the sample,
we observe a negative correlation between this ratio and the mid-infrared SED spectral index.  
This evidence describes a more complicated polar CO$_2$ environment than indicated by previous studies.  
Comparison of the SED spectral index with the ice temperatures derived from the 15.2\mic fits
confirms the presence of cold ices in more deeply embedded objects and warmer, annealed ices in more evolved ones.

Comparison with molecular cloud sight-lines show that our fits are consistent with a scenario
in which H$_2$O-rich grain mantles with CO$_2$ and CH$_3$OH relative abundances of $\sim$20 and $\lesssim$10\%, respectively, as well as traces of other species, 
form in the cold molecular clouds and are
subsequently heated by young protostars.  This thermal processing sublimates the most volatile species while inducing molecular interactions and migrations
that become spectroscopically detectable.  The wide range of both molecular abundances and thermal processing levels observed within our sample indicates that
young stars have a strong effect on their surrounding environments.  

\begin{acknowledgements}
We are grateful to Edwin Bergin and Claudia Knez for providing us with the field star spectra, to Melissa McClure for assistance with data reduction issues, and
to Adwin Boogert for stimulating discussion.
Extensive use was made of Leiden University's Sackler Laboratory ice analog databases and of the NASA ADS Abstract Service.
This work is based on observations made with the Spitzer Space
Telescope, which is operated by the Jet Propulsion Laboratory,
California Institute of Technology under NASA contract 1407. Support
for this work was provided by NASA through Contract Number 1257184
issued by JPL/Caltech.
E.F. acknowledges support from a NASA Postdoctoral Program Fellowship.
\end{acknowledgements}

\bibliographystyle{aa}
\bibliography{gaillib}

\begin{deluxetable}{l c c c} \tablewidth{0 pt} \tabletypesize{\small} 
\tablecaption{Overview of the Spitzer IRS observations of our sample of Class I/II objects.  The first column gives the object name, and the second to fourth columns
give the observations' AOR key, date, and IRS modules/mode used (SL: short-low, SH: short-high, LL: long-low, LH: long-high).} 
\tablehead{\colhead{Object} & \colhead{AOR} & \colhead{Obs. Date (2004)} & \colhead{IRS Modules/Mode}}
\startdata
IRAS 04016+2610 &3528960 & 2/07 &SL,SH,LH Map \\
IRAS 04108+2803A &3539712 & 2/06 &SL,LL Stare \\ 
IRAS 04108+2803B &3529472 & 2/06 &SL,SH,LH Map \\ 
IRAS 04154+2823 &3534336 & 3/03 &SL,LL Map \\
IRAS 04169+2702 &3534848 & 3/04 &SL,LL Map \\
IRAS 04181+2654A &3546112 & 2/29 &SL,LL Stare \\ 
IRAS 04181+2654B &3545856 & 2/07 &SL,LL Stare \\ 
IRAS 04239+2436 &3530752 & 2/08 &SL,SH,LH Map \\
IRAS 04295+2251 &3537408 & 2/27 &SL,LL Map \\
IRAS 04361+2547 &3533056 & 2/29 &SL,SH,LH Map \\ 
IRAS 04365+2535 &3533312 & 2/27 &SL,SH,LH Map \\
IRAS 04381+2540 &3538944 & 2/27 &SL,LL Map \\ 
DG Tau B &3540992 & 2/08 &SL,SH,LH Stare \\
HL Tau &3531776 & 3/04 &SL,SH,LH Map \\
IC 2087 IR &3533312 & 2/27 &SL,SH,LH Map \\
L1551 IRS 5 &3531776 & 3/04 &SL,SH,LH Map \\ 
\enddata
\label{tab:obs}
\end{deluxetable}
\clearpage

\begin{deluxetable}{c c c c c} \tabletypesize{\scriptsize} \tablewidth{0 pt}
\tablecaption{Absorption features identified in our YSO spectra.} 
\tablehead{\colhead{Wavelength ($\mu$m)} & \colhead{Carrier(s)} & \colhead{Mode} & \colhead{{\it A} (10$^{-17}$cm molec$^{-1}$)} & \colhead{Reference}}
\startdata
Ice Features & & & & \\
\hline
5.81 & H$_2$CO & C=O stretching & 0.96 & 1 \\
5.85 & HCOOH & C=O stretching & 6.7 & 2 \\
6.0 & H$_2$O & H-O-H bending & 1.2-1.3\tablenotemark{a} & 3 \\
6.8 & CH$_3$OH?, NH$_4^+$?, organics & O-H bending, C-H deforming & $\cdots$ & \\
7.58 & SO$_2$ & S-O stretching & 3.4 & 4 \\
7.7 & CH$_4$ & C-H deforming & 0.73 & 4 \\
15.2 & CO$_2$ & O-C-O bending & 1.5\tablenotemark{b} & 3 \\
\hline
Other Features & & & & \\
\hline
9.7 & Silicates & Si-O stretching & $\cdots$ & \\
14.97 & CO$_2$ & gas-phase $\nu_2$ bend & $\cdots$ & 5 \\
18-20 & Silicates & O-Si-O bending & $\cdots$ & \\
\enddata
\tablenotetext{a}{{\it A} depends on temperature; range represents 10-120K}
\tablenotetext{b}{{\it A} for H$_2$O:CO$_2$ mix, see Section~\ref{sec:column}}
\tablerefs{(1) \citet{Schutte_93_h2co}, (2) \citet{Marechal_87_hcooh}, (3) \citet{GSG_95_labsim}, (4) \citet{Boogert_97_labch4}, (5) \citet{vDHG_96_co2gasphase}}
 \label{tab:features}
\end{deluxetable}
\clearpage

\begin{deluxetable}{l c c c c c c} \tablewidth{0 pt} \tabletypesize{\small} 
\tablecaption{Characteristics of our sample of Taurus Class I/II objects.  The first column gives the object name, the second and third columns
give the 6.0\mic and 15.2\mic peak optical depths, and the fourth column gives the SED spectral index $\alpha$ (7-30 $\mu m$; Equation~\ref{eqn:alpha}).} 
\tablehead{\colhead{Object} & \colhead{$\tau_{6.0\mu m}$} & \colhead{$\tau_{15.2\mu m}$} & \colhead{SED $\alpha$}}
\startdata
IRAS 04016+2610 & 0.38($\pm$0.01) & 0.73($\pm$0.01) & 0.43($\pm$0.02) \\
IRAS 04108+2803A & 0.19($\pm$0.01)& 0.19($\pm$0.01) & 0.29($\pm$0.10) \\
IRAS 04108+2803B & 0.26($\pm$0.01)& 0.43($\pm$0.01) & 0.21($\pm$0.07) \\
IRAS 04154+2823 & 0.08($\pm$0.01) & 0.16($\pm$0.01) & -0.35($\pm$0.04) \\
IRAS 04169+2702 & 0.25($\pm$0.01) & 0.31($\pm$0.01) & 0.98($\pm$0.02) \\
IRAS 04181+2654A & 0.2($\pm$0.01) & 0.3($\pm$0.01) & 0.08($\pm$0.02) \\
IRAS 04181+2654B & 0.32($\pm$0.03) & 0.44($\pm$0.01) & 0.51($\pm$0.03) \\
IRAS 04239+2436 & 0.25($\pm$0.01) & 0.44($\pm$0.01) & 0.41($\pm$0.05) \\
IRAS 04295+2251 & 0.2($\pm$0.02)& 0.18($\pm$0.01) & 0.47($\pm$0.02) \\
IRAS 04361+2547 & 0.45($\pm$0.01) & 0.27($\pm$0.03) & 2.22($\pm$0.02) \\
IRAS 04365+2535 & 0.33($\pm$0.01) & 0.63($\pm$0.01) & 0.54($\pm$0.04) \\
IRAS 04381+2540 & 0.44($\pm$0.01) & 0.62($\pm$0.01) & 0.61($\pm$0.03) \\
DG Tau B & 0.34($\pm$0.04) & 0.3($\pm$0.01) & 0.38($\pm$0.07) \\
HL Tau & 0.12($\pm$0.01) & 0.12($\pm$0.01) & 0.34($\pm$0.02) \\
IC 2087 IR & 0.14($\pm$0.01) & 0.33($\pm$0.01) & -0.89$\pm$0.01) \\
L1551 IRS 5 & 0.54($\pm$0.01) & 0.58($\pm$0.03) & 1.41($\pm$0.02) \\
\enddata
\label{tab:obj}
\end{deluxetable}
\clearpage

\begin{deluxetable}{l c c} \tabletypesize{\small} \tablewidth{0 pt}
\tablecaption{Laboratory spectra used in 6\mic fits.} 
\tablehead{\colhead{Composition} & \colhead{Temperature (K)} & \colhead{Radiation\tablenotemark{a}}}
\startdata 
Pure H$_2$O & 10 & $\cdots$ \\
Pure H$_2$O & 30 & $\cdots$ \\
Pure H$_2$O & 50 & $\cdots$ \\
Pure H$_2$O & 120 & $\cdots$ \\
Pure H$_2$CO & 10 & $\cdots$ \\
Pure HCOOH & 10 & $\cdots$ \\
H$_2$O:CH$_4$\tablenotemark{b} 1:0.33 & 50 & 1 hr \\
H$_2$O:NH$_3$ 1:0.2 & 10 & 1 hr \\
\enddata
\tablenotetext{a}{The equivalent of 10$^{15}$cm$^{-2}$s$^{-1}$ photons (E$_\gamma>$ 6 eV) for the noted length of time}
\tablenotetext{b}{CH$_4$ does not have a resonance at 6$\mu$m, but its presence in a water ice matrix affects the shape of the feature.}
\tablerefs{All spectra from Leiden University's Sackler Laboratory ice analog databases.}
\label{tab:6um_components}
\end{deluxetable}
\clearpage

\begin{deluxetable}{l c c c c c c c c} \rotate
\tablewidth{0 pt}
\tabletypesize{\footnotesize}
\tablecaption{Derived solid-state molecular abundances in our YSO sample.  The first column gives the object name, and the second through eighth columns
give the column densities of H$_2$O, CO$_2$, H$_2$CO, HCOOH, NH$_3$, SO$_2$, CH$_4$, and CH$_3$OH.}
\tablehead{\colhead{Object} & \colhead{$N$(H$_2$O)\tablenotemark{a}} & \colhead{$N$(CO$_2$)\tablenotemark{b}} & \colhead{$N$(H$_2$CO)} & \colhead{$N$(HCOOH)\tablenotemark{c}} & \colhead{$N$(NH$_3$)} & \colhead{$N$(SO$_2$)} & \colhead{$N$(CH$_4$)} & \colhead{$N$(CH$_3$OH)\tablenotemark{d}} 
\\ \colhead{} & \colhead{10$^{18}$cm$^{-2}$} & \colhead{\% H$_2$O} & \colhead{\% H$_2$O} & \colhead{\% H$_2$O} & \colhead{\% H$_2$O} & \colhead{\% H$_2$O} & \colhead{\% H$_2$O} & \colhead{\% H$_2$O}}
\startdata
IRAS 04016+2610 & 5.91 $\pm$ 0.2 & 17.4 $\pm$ 1.4 & 1.9 & 1.8 & $<$12.1 & $<$0.1  &  0.4 $\pm$ 0.4 &  8.3 $\pm$ 1.4 \\
IRAS 04108+2803A & 2.90 $\pm$ 0.3 & 9.5 $\pm$ 2.4 & 0.3 & 0.6 & $<$9.0 &  1.7 $\pm$ 1.1  &  8.0 $\pm$ 2.6 & $\cdots$  \\
IRAS 04108+2803B & 4.44 $\pm$ 0.2 & 11.1 $\pm$ 1.5 & 2.2 & 0.8 & $<$9.0 & $<$0.7\tablenotemark{e} &  5.6 $\pm$ 1.2 &  6.3 $\pm$ 1.5 \\
IRAS 04154+2823 & 1.07 $\pm$ 0.2 & 19.3 $\pm$ 7.9 & 0.1 & 0.0 & $<$28.4 & $<$0.5  &  23.2 $\pm$ 9.1 & $\cdots$  \\
IRAS 04169+2702 & 4.33 $\pm$ 0.2 & 10.1 $\pm$ 1.5 & 2.4 & 1.0 & $<$9.0 & $<$0.5\tablenotemark{e} &  3.9 $\pm$ 1.2 & $\cdots$  \\
IRAS 04181+2654A & 3.63 $\pm$ 0.2 & 11.4 $\pm$ 1.8 & 3.6 & 0.2 & $<$9.0 & $<$0.1\tablenotemark{e} &  3.1 $\pm$ 1.4 & $\cdots$  \\
IRAS 04181+2654B & 4.18 $\pm$ 0.4 & 15.6 $\pm$ 2.3 & 7.1 & 0.7 & $<$25.6 & $<$0.3  &  4.5 $\pm$ 2.0 & $\cdots$  \\
IRAS 04239+2436 & 4.44 $\pm$ 0.2 & 12.6 $\pm$ 1.5 & 2.4 & 0.0 & $<$12.5 &  0.6 $\pm$ 0.2  & $<$0.5 &  3.8 $\pm$ 1.5 \\
IRAS 04295+2251 & 2.82 $\pm$ 0.3 & 9.4 $\pm$ 2.3 & 1.7 & 0.4 & $<$15.3 & $<$1.0\tablenotemark{e} &  2.4 $\pm$ 1.8 & $\cdots$  \\
IRAS 04361+2547 & 7.38 $\pm$ 0.3 & 4.9 $\pm$ 1.5 & 0.0 & 0.9 & $<$9.0 & $<$0.1  &  4.3 $\pm$ 0.7 &  1.7 $\pm$ 1.5 \\
IRAS 04365+2535 & 5.31 $\pm$ 0.2 & 15.2 $\pm$ 1.6 & 1.8 & 0.5 & $<$13.0 & $<$0.1  &  0.8 $\pm$ 0.7 &  8.5 $\pm$ 1.6 \\
IRAS 04381+2540 & 7.92 $\pm$ 0.2 & 11.8 $\pm$ 0.8 & 0.8 & 0.5 & $<$9.2 &  0.2 $\pm$ 0.1  &  0.3 $\pm$ 0.3 & $\cdots$  \\
DGTauB & 5.65 $\pm$ 0.6 & 7.5 $\pm$ 1.5 & 0.0 & 0.5 & $<$17.7 &  1.2 $\pm$ 0.6  & $<$0.4 &  3.5 $\pm$ 1.5 \\
HLTau & 1.91 $\pm$ 0.2 & 7.3 $\pm$ 2.8 & 5.9 & 0.7 & $<$11.0 &  1.0 $\pm$ 0.5  & $<$1.2 &  4.0 $\pm$ 2.8 \\
IC2087IR & 2.13 $\pm$ 0.2 & 19.5 $\pm$ 4.5 & 0.8 & 0.1 & $<$26.7 & $<$0.2  &  1.8 $\pm$ 2.2 &  8.7 $\pm$ 4.5 \\
L1551IRS5 & 10.90 $\pm$ 0.2 & 8.0 $\pm$ 1.3 & 0.0 & 0.0 & $<$9.0 & $<$0.0  &  1.7 $\pm$ 0.4 &  3.2 $\pm$ 1.3 \\
\enddata
\tablecomments{$N$(H$_2$CO) and $N$(HCOOH) have typical absolute uncertainties of $\sim$10\%.}
\tablenotetext{a}{Uncertainties listed are statistical, and derived from the residual differences between the data and our models. There are potential systematic errors as well, as described in Section~\ref{sec:h2o}.}
\tablenotetext{b}{Using band strength of H$_2$O:CO$_2$ mixture; see Section~\ref{sec:chemdiscuss}.}
\tablenotetext{c}{Based on 5.8\mic feature alone.}
\tablenotetext{d}{$N$(CH$_3$OH) intermixed with CO$_2$, from high-resolution 15.2\mic feature fits.}
\tablenotetext{e}{A feature was detected in the SO$_2$ wavelength range but with an overly-narrow width; value given is an upper limit.}
\label{tab:abundances}
\end{deluxetable}
\clearpage

\begin{deluxetable}{l c c c} \tabletypesize{\small} \tablewidth{0 pt}
\tablecaption{Laboratory spectra used in 15\mic fits.} 
\tablehead{\colhead{Composition} & \colhead{Temperature (K)} & \colhead{Radiation\tablenotemark{a}} & \colhead{Reference}}
\startdata 
H$_2$O:CH$_3$OH:CO$_2$ 0.9:0.3:1 & 98 & $\cdots$ & 1 \\
H$_2$O:CH$_3$OH:CO$_2$ 1:1:1 & 130 & $\cdots$ & 1 \\
CO:H$_2$O 10:1 & 30 & 1 hr & 2 \\
H$_2$O:CO$_2$ 10:1 & 10 & $\cdots$ & 3 \\
Pure CO$_2$ & 10 & $\cdots$ & 3 \\
\enddata
\tablenotetext{a}{The equivalent of 10$^{15}$cm$^{-2}$s$^{-1}$ photons (E$_\gamma>$ 6 eV) for the noted length of time}
\tablerefs{(1) \citet{EKS_99_laboratory}, (2) \citet{GSG_95_labsim}, (3) \citet{EBG_96_coco2iso}}
\label{tab:15um_components}
\end{deluxetable}
\clearpage

\begin{deluxetable}{l c c c c c c c c} \tablewidth{0 pt} \tabletypesize{\small}
\tablewidth{0 pt}
\tablecaption{Percentage and column density $N$ (10$^{17}$cm$^{-2}$) of CO$_2$ ice present in each stage of processing along the 
line of sight: cold, warm, and segregated (due to rounding, percentages may not sum to exactly 100\%).} 
\tablehead{ & \multicolumn{2}{c}{Cold} & & \multicolumn{2}{c}{Warm} & & \multicolumn{2}{c}{Segregated} \\ \cline{2-3} \cline{5-6} \cline{8-9}
\colhead{Object} & \colhead{\% Total} & \colhead{$N$} & & \colhead{\% Total} & \colhead{$N$} & & \colhead{\% Total} & \colhead{$N$}}
\tablecolumns{7}
\startdata
IRAS 04016+2610 & 3 & 0.28 & & 71 & 7.3 & & 26 & 2.6 \\
IRAS 04108+2803B & 33 & 1.6 & & 30 & 1.5 & & 37 & 1.8 \\ 
IRAS 04239+2436 & 10 & 0.57 & & 87 & 4.9 & & 2 & 0.14 \\ 
IRAS 04361+2547 & 62 & 2.2 & & 12 & 0.45 & & 26 & 0.95 \\ 
IRAS 04365+2535 & 2 & 0.17 & & 61 & 4.9 & & 37 & 2.9 \\ 
DG Tau B & 1 & 0.05 & & 76 & 3.2 & & 23 & 1.0 \\ 
HL Tau & 32 & 0.45 & & 32 & 0.44 & & 36 & 0.51 \\ 
IC 2087 IR & 9 & 0.35 & & 68 & 2.8 & & 23 & 0.95 \\ 
L1551 IRS 5 & 22 & 1.9 & & 54 & 4.6 & & 24 & 2.1 \\ 
\enddata
\label{tab:procseg}
\end{deluxetable}
\clearpage

\begin{deluxetable}{l c c c c c c c c} \tablewidth{0 pt} \tabletypesize{\footnotesize}
\tablewidth{0 pt}
\tablecaption{Molecular cloud and Class 0 abundances.}
\tablehead{ & \colhead{$N$(H$_2$O) (10$^{18}$cm$^{-2}$)} & \colhead{$N$(H$_2$O) (10$^{18}$cm$^{-2}$)} & \colhead{$N$(CO$_2$)/$N$(H$_2$O)\tablenotemark{a}} & \colhead{$N$(CO$_2$)/$N$(H$_2$O)} \\
\colhead{Source} & \colhead{[This work]} & \colhead{[Lit., (Refs)]} & \colhead{[This work]} & \colhead{[Lit., (Refs)]}}
\startdata
CK 2 [Serpens] & 2.9 & 3.5 (1) & 0.29 & 0.33 (2) \\
Elias 13 [Taurus] & 0.95 & 1.0 (3) & 0.14 & 0.15,0.19 (2,5) \\
Elias 16 [Taurus] & 2.2 & 2.5 (4) & 0.2 & 0.18,0.23 (3,5) \\
\hline
CepE-MM & $<$29.7\tablenotemark{b} & $\cdots$ & $>$0.09 & $\cdots$ \\
L1448 IRS2 & $<$28.6\tablenotemark{b} & $\cdots$ & $>$0.1 & $\cdots$ \\
W3(OH) + W3 & $\cdots$ & $\cdots$ & $\cdots$ & $\cdots$ \\
\enddata
\tablenotetext{a}{$N$(CO$_2$) calculated using the band strength $A$ of pure CO$_2$ in order to facilitate comparison with published values.}
\tablenotetext{b}{Calculated from a scaled pure H$_2$O spectrum.}
\tablerefs{(1) \citet{Eiroa_89_serpensice}, (2) \citet{KBP_05_backgroundstars}, (3) \citet{Whittet_88_extinctionlaw}, (4) \citet{CAK_95_solidcotaurus}, (5) 
\citet{Nummelin_01_co2inlowmass}}
\label{tab:molcloud_abun}
\end{deluxetable}
\clearpage

\clearpage

\clearpage

\clearpage

\clearpage

\clearpage

\clearpage

\clearpage

\clearpage

\clearpage

\clearpage

\clearpage

\clearpage

\clearpage

\clearpage

\clearpage

\clearpage

\clearpage

\clearpage

\clearpage

\clearpage

\end{document}